\documentclass[twocolumn,tighten]{aastex631}
\shorttitle{A New Flare-Mode Transitional Millisecond Pulsar}
\shortauthors{Strader \etal~}
\def\etal{{et al.}}

\def\hub{\ifmmode H_\circ\else H$_\circ$\fi}
\usepackage{subfigure}
\usepackage{amsmath}
\usepackage{graphics}

\def\ltsima{$\; \buildrel < \over \sim \;$}
\def\simlt{\lower.5ex\hbox{\ltsima}} 
\def\gtsima{$\; \buildrel > \over \sim \;$}
\def\simgt{\lower.5ex\hbox{\gtsima}} 
\def\arcsec{\hbox{$^{\prime\prime}$}}

\newcommand\T{\rule{0pt}{2.6ex}}
\newcommand\B{\rule[-1.2ex]{0pt}{0pt}}

\begin{document}

\title{Multiwavelength evidence for a new flare-mode transitional millisecond pulsar}
\author{Jay Strader}
\affiliation{Center for Data Intensive and Time Domain Astronomy, Department of Physics and Astronomy, Michigan State University, East Lansing, MI 48824, USA}
\email{straderj@msu.edu}
\author{Samuel J.~Swihart}
\affiliation{National Research Council Research Associate, National Academy of Sciences, Washington, DC 20001, USA,
resident at Naval Research Laboratory, Washington, DC 20375, USA}
\author{Ryan Urquhart}
\affiliation{Center for Data Intensive and Time Domain Astronomy, Department of Physics and Astronomy, Michigan State University, East Lansing, MI 48824, USA}
\author{Laura Chomiuk}
\affiliation{Center for Data Intensive and Time Domain Astronomy, Department of Physics and Astronomy, Michigan State University, East Lansing, MI 48824, USA}
\author{Elias Aydi}
\affiliation{Center for Data Intensive and Time Domain Astronomy, Department of Physics and Astronomy, Michigan State University, East Lansing, MI 48824, USA}
\author{Arash Bahramian}
\affiliation{International Centre for Radio Astronomy Research, Curtin University, GPO Box U1987, Perth, WA 6845, Australia}
\author{Adam Kawash}
\affiliation{Center for Data Intensive and Time Domain Astronomy, Department of Physics and Astronomy, Michigan State University, East Lansing, MI 48824, USA}
\author{Kirill V. Sokolovsky}
\affiliation{Center for Data Intensive and Time Domain Astronomy, Department of Physics and Astronomy, Michigan State University, East Lansing, MI 48824, USA}
\affiliation{Sternberg Astronomical Institute, Moscow State University, Universitetskii pr. 13, 119992 Moscow, Russia}
\author{Evangelia Tremou}
\affiliation{LESIA, Observatoire de Paris, CNRS, PSL, SU/UPD, Meudon, France}
\author{Andrej Udalski}
\affiliation{Astronomical Observatory, University of Warsaw, AI. Ujazdowskie 4, 00-478 Warszawa, Poland}

\begin{abstract}

We report the discovery of a new low-mass X-ray binary near the center of the unassociated \emph{Fermi} GeV $\gamma$-ray source 4FGL J0540.0--7552. The source shows the persistent presence of an optical accretion disk and exhibits extreme X-ray and optical variability. It also has an X-ray spectrum well-fit by a hard power law with $\Gamma = 1.8$ and a high ratio of X-ray to $\gamma$-ray flux. Together, these properties are consistent with the classification of the binary as a transitional millisecond pulsar (tMSP) in the sub-luminous disk state. Uniquely among the candidate tMSPs, 4FGL J0540.0--7552 shows consistent optical, X-ray, and $\gamma$-ray evidence for having undergone a state change, becoming substantially brighter in the optical and X-rays and fainter in GeV $\gamma$-rays sometime in mid-2013. In its current sub-luminous disk state, and like one other candidate tMSP in the Galactic field,  4FGL J0540.0--7552 appears to always be in an X-ray ``flare mode", indicating that this could be common phenomenology for tMSPs. 

\end{abstract}
 
\section{Introduction}

Millisecond pulsars (MSPs) form when a neutron star in a low or intermediate-mass X-ray binary accretes matter from its companion star, spinning up to a fast rotation rate \citep{Alpar82,Bhattacharya91}. Most MSPs in the Galactic field are fully recycled, with accretion ceased permanently, and have low-mass helium white dwarf companion stars \citep{Tauris99}.

An unexpected finding from the all-sky GeV $\gamma$-ray survey of the \emph{Fermi} Large Area Telescope (LAT) is that MSPs channel a substantial fraction of their spindown energy into $\gamma$-rays, making them nearly ubiquitous $\gamma$-ray emitters \citep{Abdo13}.  Radio, X-ray, and optical follow-up of \emph{Fermi}-LAT sources have revealed a large population of short orbital period ``spider" MSP binaries with hydrogen-rich, low mass (redback; $\sim 0.1$--$0.5 M_{\odot}$), or ultra light (black widow; $\lesssim 0.05 M_{\odot}$) secondaries being ablated by the MSP \citep{Ray12,Roberts13}. These binaries had mostly been hidden from all-sky radio surveys by extensive eclipses, and can be challenging to detect as pulsars even with deep pointed observations.

These discoveries of spider MSPs show that the population of binary MSPs is richer than previously thought. They have also helped to elucidate the relationship between MSPs and X-ray binaries. In particular, three redbacks have been identified as transitional MSPs (tMSPs) that actively switch between an X-ray faint radio pulsar state (with $L_X \sim 10^{31}$--10$^{32}$ erg s$^{-1}$ over  0.5--10 keV) and a ``sub-luminous" accretion disk state (with $L_X \sim 10^{33}$--10$^{34}$ erg s$^{-1}$) on timescales of days to years \citep{Archibald09,Papitto13,Bassa14,Roy15}. 

In this sub-luminous disk state (named because typical persistent or outbursting low-mass neutron star X-ray binaries have $L_X \gtrsim 10^{36}$ erg s$^{-1}$; \citealt{Paradijs98}), the radio pulsar emission is not detected, but rapid X-ray and optical variability is commonly observed (e.g., \citealt{deMartino13,Linares14,Bogdanov15}). This variability is attributed in some manner to the interaction between the neutron star and inner disk (e.g., \citealt{Papitto15,Campana19,Veledina19}). tMSPs occupy a liminal space between rotation-powered and accretion-powered neutron star binaries, and offer insights into pulsar recycling, accretion physics at low accretion rates, and the formation of compact binaries (see the recent review of \citealt{Papitto20}).

While MSPs are bright in $\gamma$-rays, no normal low-mass X-ray binaries with $L_X < 10^{36}$ erg s$^{-1}$ show persistent GeV $\gamma$-ray emission. Hence it is notable that 
both of the confirmed field tMSPs show $\gamma$-ray emission in the sub-luminous disk state \citep{Stappers14,Johnson15}. This has led newly discovered low-mass X-ray binaries with positions compatible with \emph{Fermi}-LAT GeV $\gamma$-ray sources to be consistently characterized as \emph{candidate} tMSPs. The three sources in this category are 3FGL J1544.6--1125, which shows the X-ray moding unique to tMSPs \citep{BH15}, 3FGL J0427.9--6704, for which the optical/X-ray and $\gamma$-ray association is proven via periodic eclipses \citep{Strader16,Kennedy20}, and the recent discovery 4FGL J0407.7--5702 \citep{Miller20}. A fourth source, CXOU J110926.4--650224, is a suspected low-mass X-ray binary that shows optical and X-ray variability similar to the tMSPs including X-ray moding \citep{CotiZelati19}. This binary was tentatively associated with an 8-year LAT source that is not present in 4FGL DR2 \citep{4FGLDR2}, so it is unclear whether this source has detectible $\gamma$-ray emission.

In globular clusters, the identification of candidate tMSPs is less straightforward, since there is often unresolved $\gamma$-ray emission from a population of normal MSPs (e.g., \citealt{Hooper16}). There is a single confirmed globular cluster tMSP, IGR J18245--2452 in M28 \citep{Papitto13,Linares14}. Candidates have been mooted in other clusters primarily based on unusual X-ray variability or flat-spectrum radio continuum emission (e.g., \citealt{Bahramian18,Bahramian20}), but these remain tentative.

Here we present the discovery of a new low-mass X-ray binary near the center of the unassociated \emph{Fermi}-LAT source 4FGL J0540.0--7552, and show that the source has properties suggesting it is a tMSP in the sub-luminous disk state. However, unlike the other candidate tMSPs, this binary shows consistent optical, X-ray, and $\gamma$-ray evidence for having undergone a state change in mid-2013, becoming substantially brighter in the optical and X-rays and fainter in GeV $\gamma$-rays.

\section{Data}

\subsection{$\gamma$-ray}

4FGL J0540.0--7552 is an unassociated source in the 10-year 4FGL DR2 catalog \citep{4FGLDR2,4FGL}. It is of high significance in this catalog ($\sim 15\sigma$) and first appeared as a LAT source in the 2-year 2FGL catalog \citep{2FGL}. In 4FGL DR2, the LAT spectrum has significant curvature: a LogParabola model is preferred over a power law at $3.3\sigma$. In both 4FGL and 4FGL DR2, 4FGL J0540.0--7552 is formally classified as a variable source. In this paper we use both the integrated (10-year) properties of 4FGL J0540.0--7552 as well as the 1-yr binned light curve, all taken from 4FGL DR2.

\subsection{X-ray}

\subsubsection{Swift and ROSAT}

After its appearance in the 2FGL catalog, the region containing 4FGL J0540.0--7552 was imaged with \emph{Swift}/XRT on three epochs from Jan to May 2012, for about 4.5 ksec of total exposure time, as part of an ongoing program to image unassociated \emph{Fermi} sources \citep{Stroh13}. We analyzed these data using the online XRT product tools \citep{Evans20}, finding two significant sources within the 4FGL DR2 68\% error ellipse (and none outside the 68\% ellipse but within the 95\% ellipse). The brighter of these sources, which we term J0540A, is more distant from the $\gamma$-ray center ($\sim 2.0$\arcmin), with about 17 net XRT counts and a 0.3--10 keV count rate of  $(4.9\pm1.2) \times 10^{-3}$ counts s$^{-1}$. The fainter X-ray source, which we call J0540B, has about 7 net counts and a count rate of $(2.0\pm0.8) \times 10^{-3}$ counts s$^{-1}$. The count rate is too low to meaningfully assess variability between the \emph{Swift} epochs. J0540B is somewhat closer to the center of the \emph{Fermi} error ellipse, separated by $\sim 1.7$\arcmin. We note that both of these sources are also listed in the second \emph{Swift}/XRT point source catalog \citep{Evans20} with parameters consistent with, but not identical to, these values. 

Owing to its location near the south ecliptic pole, this region had relatively deep \emph{ROSAT} X-ray imaging from 1990 to 1991, totaling 5.1 ksec with the Position Sensitive Proportional Counter (PSPC) instrument. We analyzed these data using the {\it Sosta} task in XIMAGE, distributed as part of {\tt HEAsoft} version 6.28 \citep{2014ascl.soft08004N}. There is no significant X-ray source at the location of J0540B, with a $3\sigma$ upper limit to the 0.1--2.4 keV count rate of $< 3.7 \times 10^{-3}$ ct s$^{-1}$.

\begin{figure*}[t!]
\includegraphics[width=6.8in]{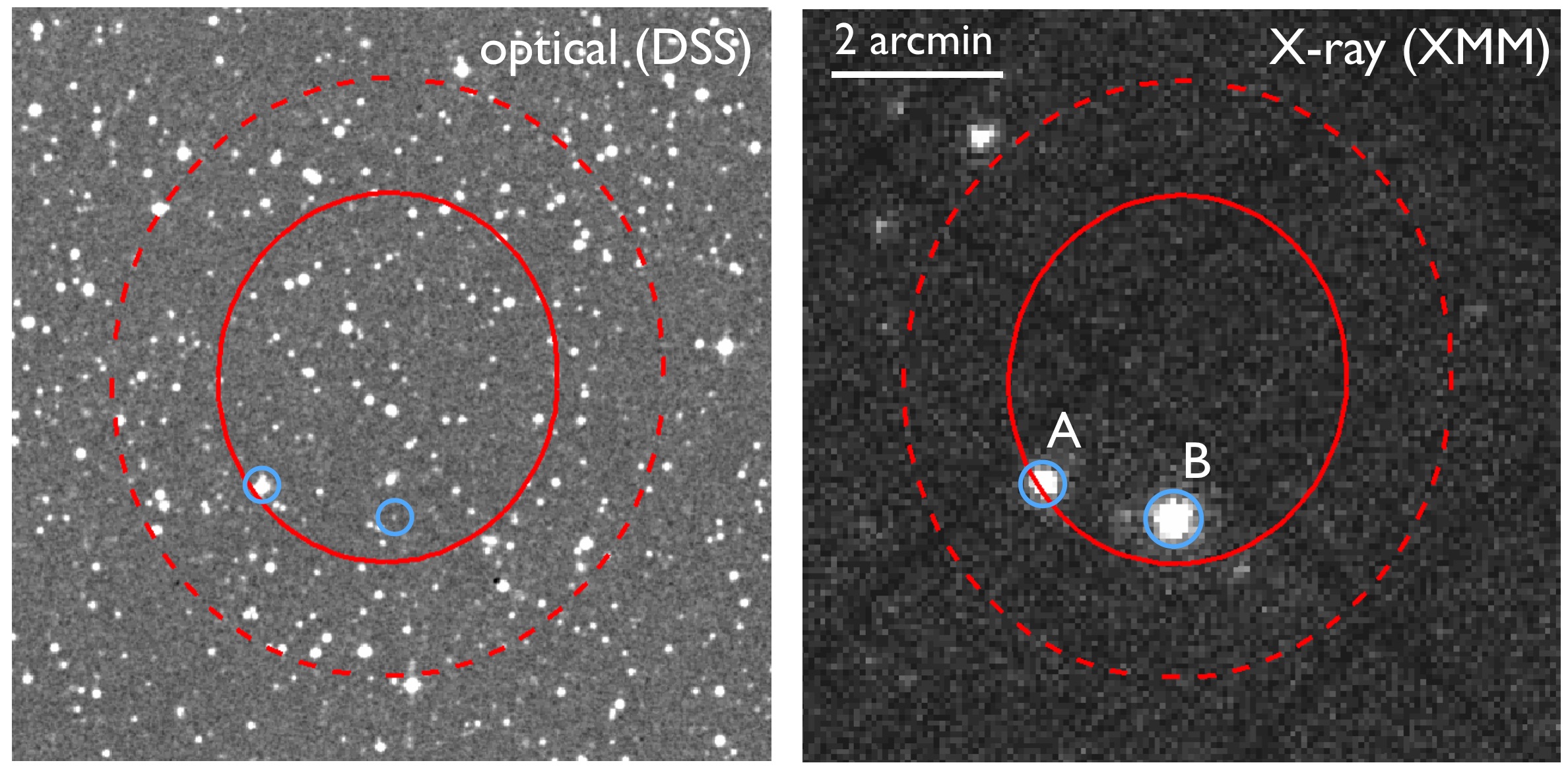}
\caption{Optical (left) and \emph{XMM} X-ray (right) images of the region around 4FGL J0540.0--7552. In each panel the \emph{Fermi}-LAT 68\% (solid red) and 95\% (dashed red) error ellipses are plotted. The two bright X-ray sources within the 68\% error ellipse are circled in pale blue: J0540A (fainter in \emph{XMM}, brighter in optical); J0540B (brighter in \emph{XMM}, fainter in optical). J0540B  is the likely counterpart to the GeV $\gamma$-ray source.}
\label{fig:finder}
\end{figure*}

\subsubsection{XMM}

On 2021 Jan 23, we used \emph{XMM-Newton} with the European Photon Imaging Camera (EPIC) to observe the region centered on 4FGL J0540.0--7552, with an on-source live exposure time of 41.8 ksec. The observation was reprocessed using standard tasks in the Science Analysis System (SAS, version 18.0.0, \citealt{2004ASPC..314..759G}). Intervals of high particle background were filtered out. For both the spectrum and light curve, we used a 30\arcsec\ circular source extraction region, and a local background region at least three times larger, taking care to avoid any chip gaps. Standard flagging criteria \verb|#XMMEA_EP| and \verb|FLAG=0| were used for pn and \verb|#XMMEA_EM| was used for MOS. Additionally, we selected patterns 0--4 and 0--12 for pn and MOS, respectively. For our timing analysis, barycentric corrections were applied using the {\it barycen} task in SAS. We used the tasks {\it evselect} and {\it lccorr} to produce background-subtracted light curves for each camera. The pn, MOS1, and MOS2 light curves were then combined into a single EPIC light curve using the {\tt FTOOLS} package \textit{lcmath} \citep{1995ASPC...77..367B}. For the spectral analysis, we used the SAS task {\it xmmselect} to extract individual pn, MOS1, and MOS2 spectra before combining them with {\it epicspeccombine}. The resulting EPIC spectrum was grouped into a minimum of 20 counts per bin so that we could use Gaussian statistics. Spectral fitting was performed using {\tt XSPEC} version 12.10.1 \citep{1996ASPC..101...17A}.

The right panel of Figure \ref{fig:finder} shows the center of the \emph{XMM}/EPIC X-ray image with the \emph{Fermi}-LAT 4FGL DR2 68\% and 95\% error ellipses of 4FGL J0540.0--7552 superposed. As in the previous \emph{Swift}/XRT data, there are two bright X-ray sources within the 68\% error ellipses: J0540A and J0540B. However, in these new \emph{XMM} data, J0540B is brighter than J0540A.

\subsection{Optical Counterparts}

The X-ray source J0540A has an unambiguous match to a bright ($G=13.8$), somewhat red ($BP-RP = 1.4$) source in \emph{Gaia} EDR3 \citep{EDR3} with an ICRS position of (R.A., Dec.) = (05:40:26.948, --75:53:53.65). There are no other optical sources within 7\arcsec. This source has a very precise EDR3 parallax, giving a distance of $925\pm10$ pc \citep{BJ21}, and as discussed further below, is a confirmed variable star \citep{Chen18}.

Similarly, a single optical source is consistent with the location of J0540B in the \emph{Swift} and \emph{XMM} X-ray data, and its association is confirmed by the optical spectroscopy discussed in Section 3.6. It is present as a faint optical source in older catalogs such as USNO-B \citep{Monet03}, but the most precise information is available in \emph{Gaia} EDR3 \citep{EDR3}, where it is listed at an ICRS position of (R.A., Dec.) = (05:40:01.892, --75:54:19.26). It has $G=20.189\pm0.023$, which is an unusually high photometric uncertainty for an isolated star at this $G$ mag, strongly suggesting it is a variable star \citep{Andrew21}. This inference is supported by the new photometry discussed below in Sections 3.3 and 3.4. It has a large proper motion of $\mu_{\rm tot} = 13.45\pm0.63$ mas yr$^{-1}$, derived from a high quality astrometric fit (EDR3 renormalized unit weight error of 1.015). There is little to be learned from its corrected parallax of $\varpi = 0.121\pm0.444$ mas \citep{Lindegren20}.

The left panel of Figure \ref{fig:finder} shows a Digitized Sky Survey image with the optical counterparts to the X-ray sources marked.

\subsubsection{UV/Optical Photometry: OGLE, ASAS-SN, SOAR, XMM}

The field containing J0540A and J0540B was observed in the $I$ filter during a subset of operations of the OGLE survey of the Magellanic Clouds \citep{Udalski15}, though 
J0540B is relatively faint. We use the photometry for J0540B, retaining  all $5\sigma$ detections from the time range in which the field was regularly observed (late 2012 to early 2015). 

For J0540A, the OGLE data are relatively sparse, so we instead use data from the ASAS-SN survey \citep{Shappee14,Kochanek17}. Focusing just on the more abundant $g$ filter, there are 
over 4700 clean ASAS-SN detections of the source from 2017 Oct to 2021 Apr. 

We obtained additional photometry of J0540B with the Goodman Spectrograph \citep{Clemens04} on the SOAR telescope on 2021 Jan 13 and 19. In the first case the photometry covered a nearly 
uninterrupted interval of 6.3 hr; in the second case the data were taken over 5 hr, but with two gaps that total about 50 min. To maximize the throughput (and hence minimize the per-image exposure time), and in lieu of the availability of a clear filter, all observations were made with a GG395 long-pass filter. The exposure time for each image was 30 s, and the readout and overhead between exposures took a median of 5.6 s.

We performed differential photometry of J0540B with respect to a set of 19 bright non-variable comparison stars in the field. Given that our filter bandpass is not too dissimilar from \emph{Gaia} $G$, we calibrated our photometry to this system.  The final number of independent SOAR photometric data points for J0540B is 1010. This photometry is listed in Table \ref{tab:soar_tab}.

Finally, during our \emph{XMM} observations on 2021 Jan 23, we obtained 10 images with the Optical/UV Monitor Telescope (OM) in imaging mode, each with exposure time 3500 or 3800 s. 
These all used the $uvm2$ filter, which has an effective wavelength of 2310 \AA. We used the SAS task \textit{ommosaic} to stack all 10 images, resulting in a  
total OM exposure time of 36.5 ksec. We performed aperture photometry on J0540B using the SAS tasks \textit{omdetect} and \textit{ommag} with standard input parameters.

\subsection{Optical Spectroscopy: SOAR and Gemini}

We obtained 14 optical spectra of J0540A with the Goodman Spectrograph on the SOAR telescope on several nights over the date range 2016 Sep 19 to 2016 Dec 31. All spectra had exposure times of 10 min, and used a 1200 l mm$^{-1}$ grating that yielded a resolution of about 1.7 \AA\ over a wavelength range 5480--6740 \AA.

We obtained 10 optical spectra of J0540B with SOAR/Goodman on 2020 Dec 20 and 2021 Jan 9. Each of the spectra had the same exposure time (25 min) and covered the same wavelength range ($\sim 3950$--7850 \AA) at a resolution of 5.6 \AA. Subsequently, we obtained six spectra on the nights of 2021 Feb 4, 5 and 6 with GMOS-S on the Gemini South telescope under program code GS-2021A-FT-101. Each of these Gemini spectra were 20 min in length, and covered a wavelength range $\sim 4600$--9150 \AA\ at a resolution of 7.1 \AA. All the SOAR and Gemini spectra were reduced and optimally extracted in the normal manner using tools in IRAF \citep{Tody86}.

\subsection{Radio}

While no new radio continuum data are presented in this paper, the region around 4FGL J0540.0--7552 was imaged by the Australia Telescope Compact Array in 2013 and 2014. No radio continuum source within the \emph{Fermi}-LAT error ellipse of this source was detected, down to a limit of $\lesssim 1.5$ mJy \citep{Schinzel17}. 

\section{Results for J0540B: The Likely Counterpart to the $\gamma$-ray Source}

The results from the X-ray and optical analysis discussed below show that J0540B is overwhelmingly likely to be associated with the GeV $\gamma$-ray source 4FGL J0540.0--7552. Hence, we discuss J0540B first, followed by a section on the properties and classification of the other candidate counterpart, J0540A. Note that most of the detailed discussion and interpretation of our results on J0540B is deferred to Section 5.

\subsection{X-ray: XMM}

J0540B is the brightest X-ray source in the \emph{XMM}/EPIC data from 2021 Jan 23 located within the 68\% (or 95\%) Fermi-LAT error ellipse. We fit the \emph{XMM} X-ray spectrum alternately using a standard absorbed power-law, an absorbed thermal disk, or a combination thereof. In all cases the absorption model was {\tt tbabs} using {\tt wilm} abundances \citep{Wilms00}. 

\begin{figure}[t]
\includegraphics[width=2.4in,angle=270]{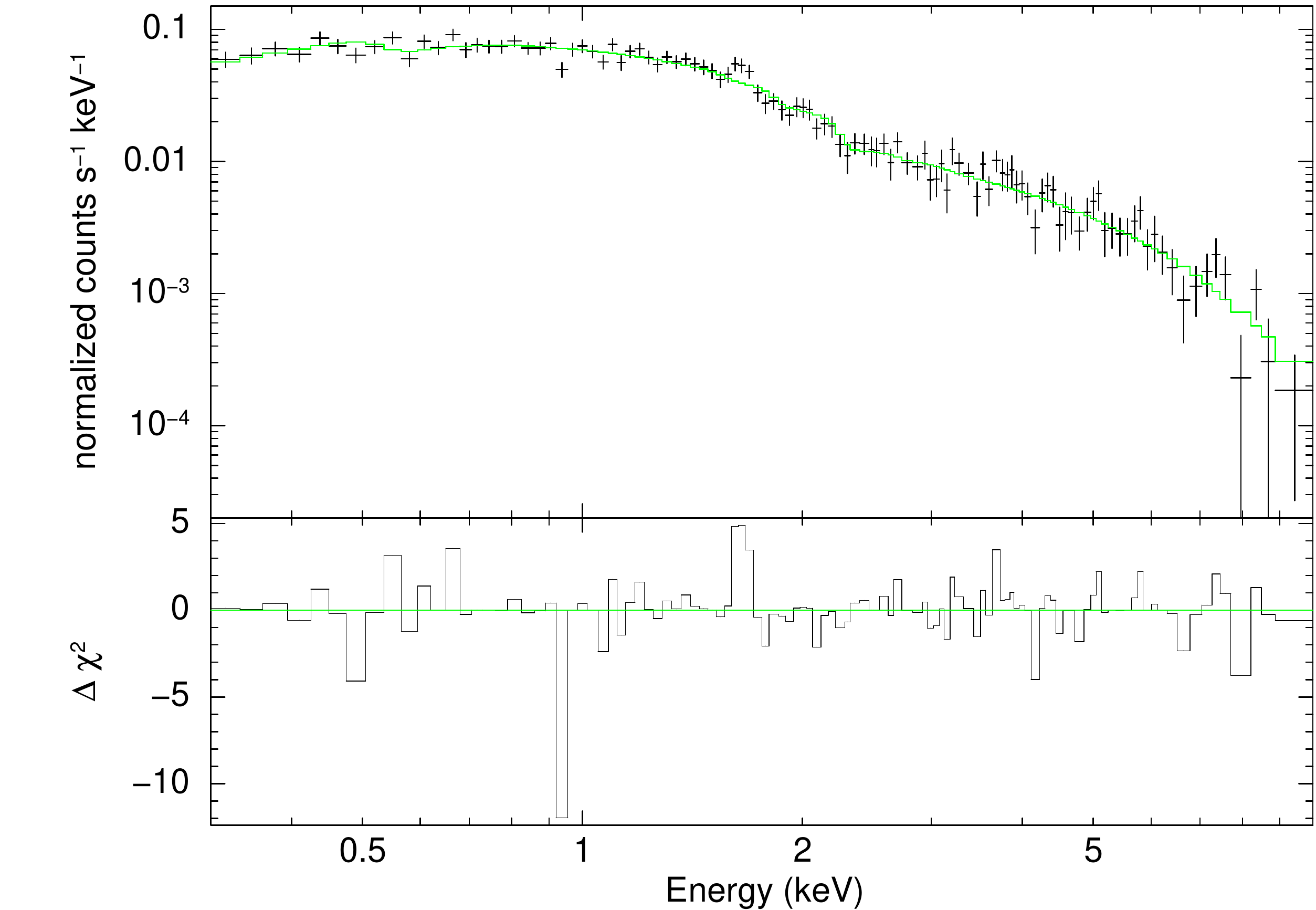}
\caption{The top panel shows the binned \emph{XMM}/EPIC X-ray spectrum of J0540B (black points with uncertainties) with the best-fitting absorbed power-law ($\Gamma = 1.78\pm0.07$) overplotted (green solid line). The bottom panel shows the residuals.}
\label{fig:xray_spec}
\end{figure}

While the thermal model is a poor fit to the data, an absorbed power-law provides an excellent fit ($\chi^2$/d.o.f. = 109.5/110). Adding a thermal component to this power law does not significantly improve the fit. The best-fit absorbed power law has photon index $\Gamma = 1.78\pm0.07$ and column density $N_H = (4.6\pm1.5) \times 10^{20}$ cm$^{-2}$, with an unabsorbed 0.5--10 keV flux of $(3.17\pm0.12) \times 10^{-13}$ erg s$^{-1}$ cm$^{-2}$. The $N_H$ value is consistent with that expected from the estimated foreground reddening of $E(B-V) = 0.07$ mag \citep{Schlafly11,Bahramian15,Guver09}. The X-ray spectrum is plotted in Figure \ref{fig:xray_spec}.

\begin{figure*}[t]
\includegraphics[width=7.0in,angle=0]{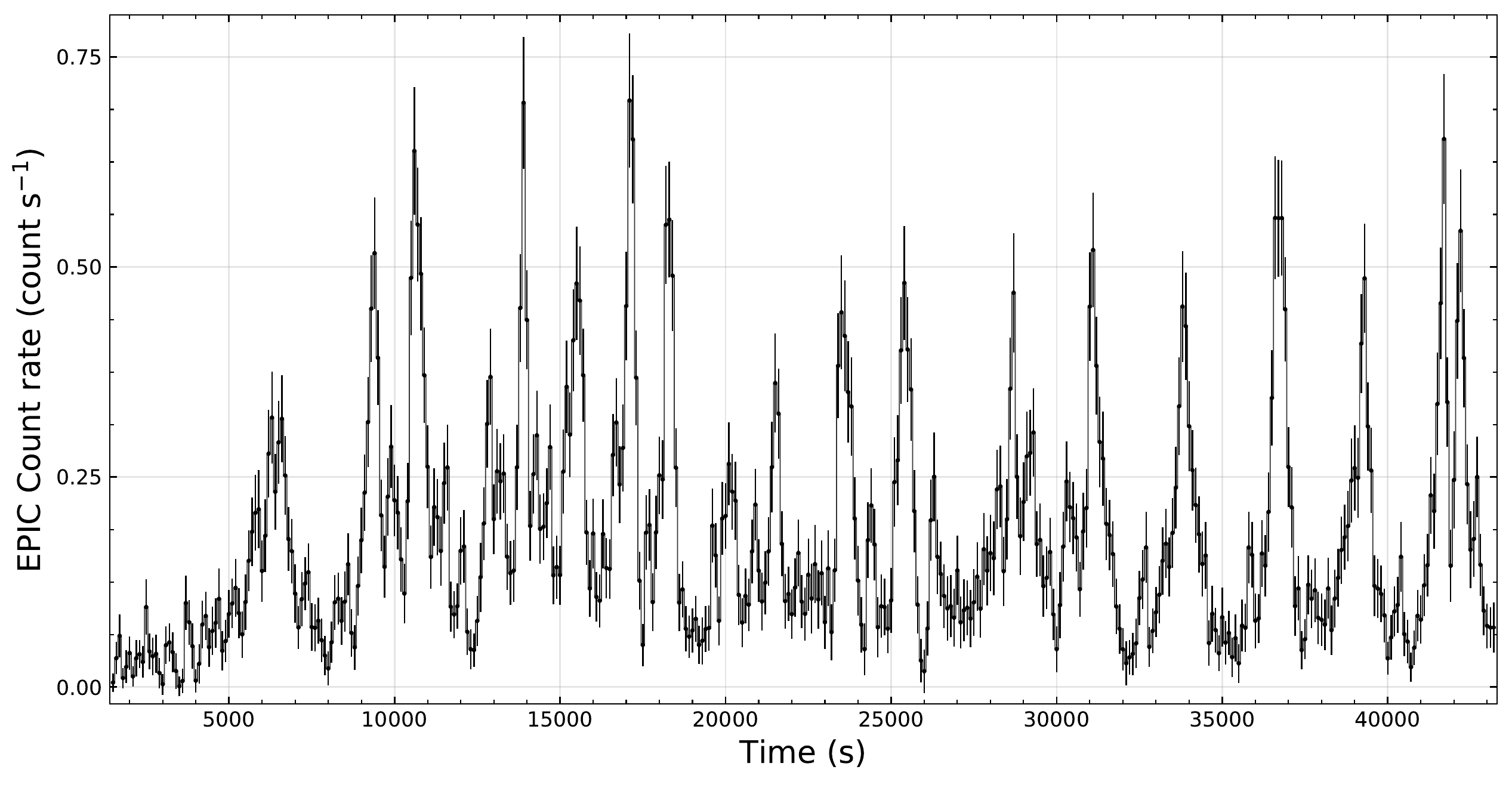}
\caption{\emph{XMM}/EPIC background-subtracted 0.2--10 keV light curve of J0540B, in 100-s bins. We identify 21 ``major" flares that peak at count rates $> 0.3$ ct s$^{-1}$, and the majority of the 42 ksec light curve is spent in a flare.}
\label{fig:xray_lc}
\end{figure*}

The background-subtracted 0.2--10 keV \emph{XMM}/EPIC X-ray light curve (Figure \ref{fig:xray_lc}) shows the same qualitative behavior for essentially the entire 42 ksec time span: repeated luminous extended flares separated by short periods of more quiescent behavior.

Owing to the modest count rate, making an accurate count of the number of flares is challenging. Just focusing on the ``major" flares, which we define as those with a peak count rate  
$> 0.3$ ct s$^{-1}$ in the 100-s binned light curve  (Figure \ref{fig:xray_lc}), there are 21---on average one every $\sim 30$ min. Around the second half of the \emph{XMM} observation the major flares appear to recur quasi-periodically, with a timescale around 45 min. Lomb-Scargle periodigrams of the entire 10-s and 100-s light curves, as implemented in {\tt R} \citep{Ruf99}, both show significant peaks around this timescale; these are also the highest-power peaks in the periodigrams. While the starting and ending times of the flares are not necessarily well-defined, the major flares seem to typically last around $\sim 1000$--1250 s, though a few last longer and some of the less luminous flares only last a few hundred sec.

We emphasize that these major flares are only a subset of the full distribution of flares. Less luminous flares are also present, but difficult to systematically catalog owing to the low count rate.
As an initial attempt, we identified ``minor" flares as those with a peak count rate $> 0.1$ ct s$^{-1}$ that are separated by at least 500 s from the peak of a major flare, finding 32 minor flares that meet these criteria. Hence these occur at least 50\% more frequently than major flares. Unlike the major flares, the minor flares do not appear to recur on any particular timescale.

The major flares are luminous: considering the 10-s binned light curve, a few of the brightest flares reached count rates $> 1.2$ ct s$^{-1}$ for $\sim 10$--30 s (equivalent to an unabsorbed 0.5--10 keV flux of $\gtrsim 2.1 \times 10^{-12}$ erg s$^{-1}$ cm$^{-2}$), but the flares more typically peaked around $\sim 0.7$--1 ct s$^{-1}$. The majority of the \emph{XMM} light curve is spent in an active flare: 67\% if the quiescent threshold is conservatively set at $< 0.1$ ct s$^{-1}$ in the 100-s binned light curve. The longest quiescent period is at the start of the light curve, for $\sim 3$--3.5 ksec before the first major flare. A typical mean quiescent count rate is about 0.06 ct s$^{-1}$, equivalent to an unabsorbed 0.5--10 keV flux of $10^{-13}$ erg s$^{-1}$ cm$^{-2}$. However, the mean flux between major flares is variable, likely due to less luminous flares during some of these periods.

We also re-fit the X-ray spectra during the flaring and quiescent periods, defining the flaring spectrum by the major flares in the 100-s binned light curve, with the starting and ending times for each flare determined with respect to the estimated local baseline. The remainder of the light curve is considered to be quiescent. While any such division is imperfect, we experimented with different choices and obtained consistent results. We find no evidence for a significant change in $\Gamma$ (flaring: $1.87\pm0.05$, quiescent: $1.79\pm0.07$) or $N_H$ (flaring: $3.3\pm1.1 \times 10^{20}$ cm$^{-2}$ , quiescent: $5.4\pm1.7 \times 10^{20}$ cm$^{-2}$). Even if this change in $N_H$---which is not significant---were taken literally, it would correspond to only a $\sim 6\%$ expected change in the observed 0.2--10 keV count rate. Hence we can rule out the possibility that the flux variations are due to changing obscuration.

\subsection{X-ray: Swift and ROSAT}

Owing to the small number of counts in the 2012 \emph{Swift}/XRT observations of J0540B, we simply adopt the \emph{XMM} spectral fit to infer the unabsorbed flux, finding a 0.5--10 keV flux of 
$(7.5\pm3.0) \times 10^{-14}$ erg s$^{-1}$ cm$^{-2}$. This flux is about a factor of 4 fainter than the mean flux inferred from the \emph{XMM} observations (with a $1\sigma$ range of a factor of 7 fainter to a factor of 3 fainter).

Since the \emph{XMM} light curve shows substantial variability, we wanted to assess whether the \emph{Swift} flux might still be consistent with that from the \emph{XMM} data. We simulated the \emph{Swift} observations by measuring the mean flux in subsets of the \emph{XMM} data with an exposure time distribution similar to that of the  \emph{Swift} data. Only about 0.1\% of the simulated observations had a flux as low as actually observed in the \emph{Swift} data. 

This is consistent with an interpretation that the mean X-ray flux of J0540B did indeed increase between 2012 May and 2021 Jan. One possibility is that the system became brighter overall in X-rays. Another possibility is that the quiescent and flaring X-ray fluxes themselves did not change, but that the rate of flares dramatically increased. The \emph{Swift}/XRT flux is compatible within $1\sigma$ with that observed during the quiescent (non-flaring) part of the \emph{XMM} light curve, which is consistent with this interpretation. We note that if the spectral behavior of the source changed over this interval, then the conversion between count rate and flux for the \emph{Swift}/XRT data will strictly not be accurate, but for essentially any reasonable spectral change the shot noise uncertainty is larger than the systematic spectral uncertainty.

The earlier \emph{ROSAT} data (taken from 1990 to 1991) also provides an informative non-detection, with the 0.1--2.4 keV $3\sigma$ upper limit of $< 3.7 \times 10^{-3}$ ct $^{-1}$ corresponding to an unabsorbed 0.5--10 keV flux limit of $< 1.0 \times 10^{-13}$ erg s$^{-1}$ cm$^{-2}$ (using the \emph{XMM} spectral fit). This upper limit is consistent with the 2012 \emph{Swift}/XRT detection, but not the brighter 2021 \emph{XMM} data. An equivalent view is that the predicted mean \emph{ROSAT}  count rate based on the \emph{XMM} observations would be $1.2 \times 10^{-2}$ ct $^{-1}$, a factor of three higher than the observed $3\sigma$ upper limit. There are quite a few cataloged \emph{ROSAT}  sources in this region of the sky fainter than this predicted value \citep{ROSAT}, implying that a source with the same mean flux as in the \emph{XMM} data should have been well-detected by \emph{ROSAT}. 

The \emph{Swift} and \emph{ROSAT} observations give two independent data points suggesting that J0540B was on average fainter in the X-rays both in the early 1990s and in 2012 compared to 2021 Jan, though whether this was due to an overall change in the X-ray flux or due instead to a difference in the occurrence of flares is uncertain.

\subsection{Optical Photometry: OGLE}

\begin{figure}[t]
\includegraphics[width=3.4in]{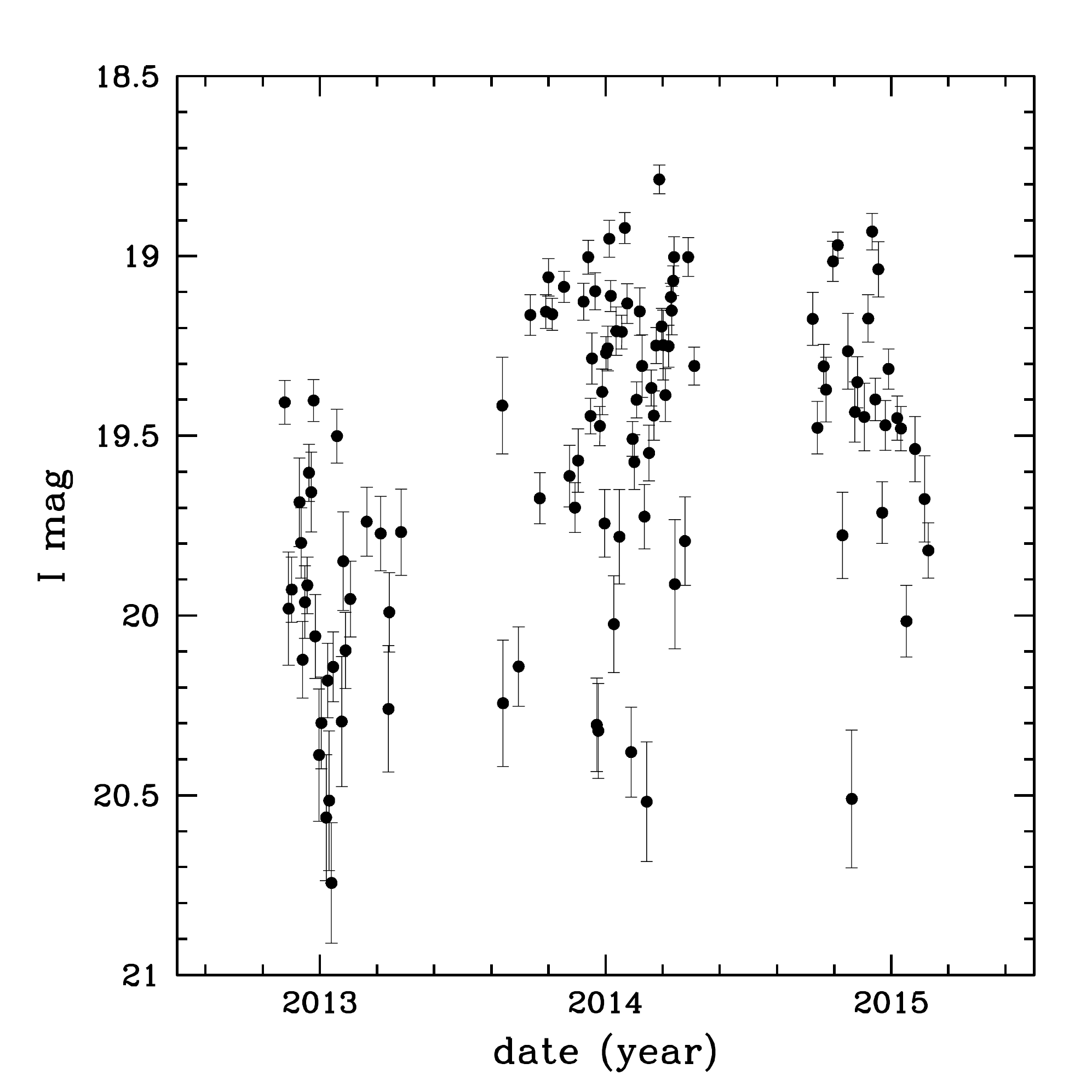}
\caption{$I$ band light curve for J0540B from OGLE, with the data spanning 2012 Nov to 2015 Feb. While the system is variable at all epochs, a $\sim 0.6$ mag increase in mean brightness
is observed between the first and second observing seasons, with the change inferred to lie in the time range 2013 Apr to 2013 Aug.}
\label{fig:ogle}
\end{figure}

The OGLE $I$ photometry offers modestly dense coverage from late 2012 to early 2015, with a data point every $\sim 4$ d in the median. There are two main conclusions that can be drawn from these data, which are plotted in Figure \ref{fig:ogle}. First, J0540B is strongly variable over the full time range of these data, with an amplitude $> 1$ mag. Second, there is strong evidence for a change in the median magnitude of the system: it is $I=19.96\pm0.07$ for the first season, brightening to $I=19.37\pm0.05$ for the next two densely sampled seasons, from mid-2013 to mid-2015 (the uncertainties in these medians were calculated via bootstrap). Owing to the seasonal gap in observations and to the large variability of the source, the precise time of the change cannot be identified, but it seems to have occurred during the seasonal gap from  2013 Apr to 2013 Aug; a date as late as 2013 Sep 11 cannot be ruled out.

To ensure that the apparent change in brightness was not due to an aphysical change in the system (e.g., a change in calibration or zeropoint), we also examined light curves for several ``check stars" in the same OGLE field that had similar brightness to J0540B. None of these stars showed any significant changes in their mean brightness over the same timespan.

There are no other obvious differences in the photometric behavior between the ``early" and ``late" measurements other than the change in mean magnitude. Formally the dispersion and range of the data are slightly larger after the 2013 change than before, but these differences are of marginal significance. In addition, OGLE becomes incomplete at $I \gtrsim 21$ mag, a range relevant to the first season given the fainter mean brightness.

Using a Lomb-Scargle periodigram, we found no evidence for a periodic signal in any seasonal subset of the data or in the whole OGLE dataset.

\subsection{Optical Photometry: SOAR}

\begin{figure}[t]
\includegraphics[width=3.4in]{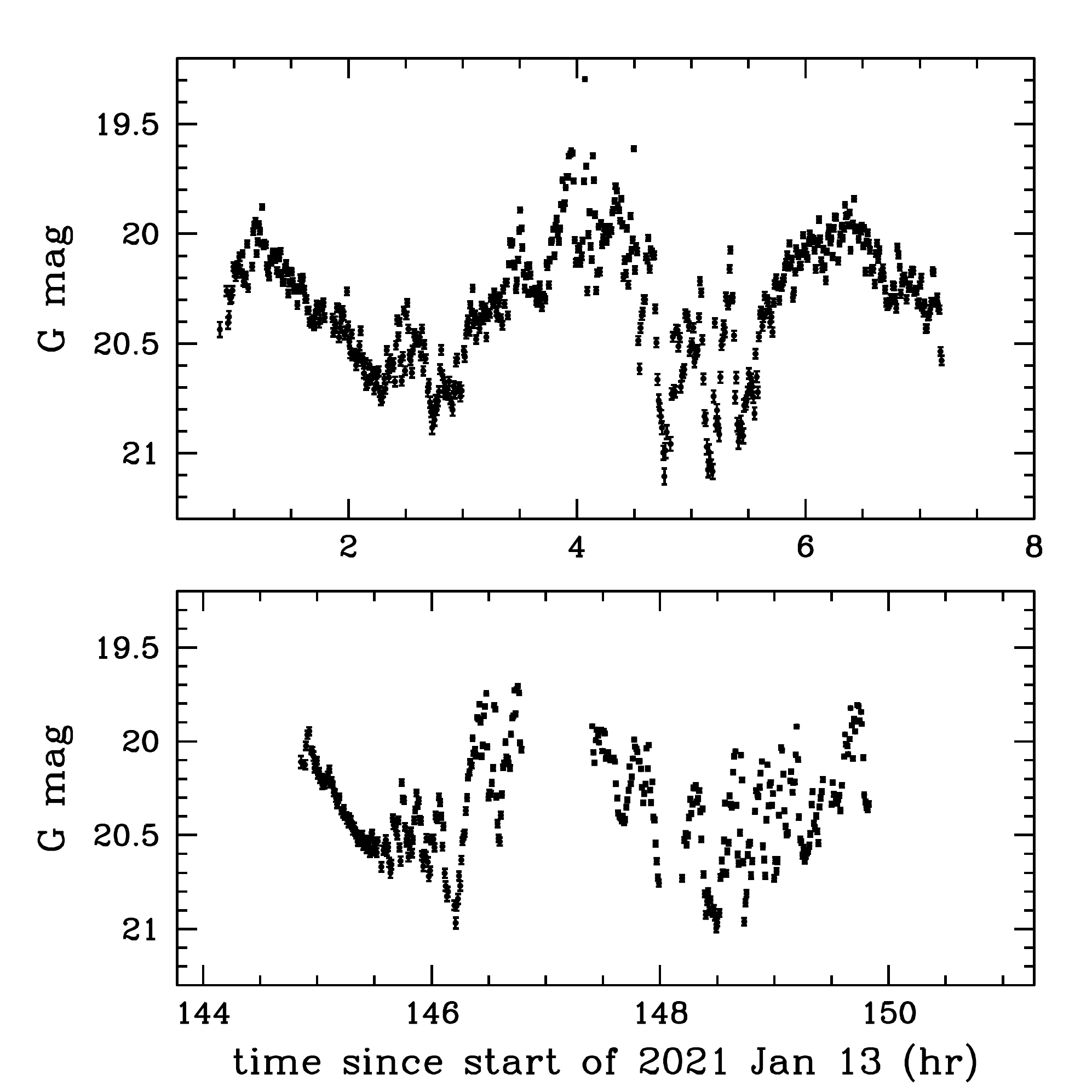}
\caption{$G$ light curve for J0540B from SOAR/Goodman, with data taken on two different nights in 2021 Jan. Extreme optical variability (amplitude of 1.8 mag), as previously
observed in the OGLE data, is also present here. Uncertainties are plotted for each data point: all the observed variations reflect intrinsic changes in the source.}
\label{fig:soar_phot}
\end{figure}

The SOAR photometry, taken in a GG395 long-pass filter (and calibrated to \emph{Gaia} $G$), offer dense sampling over two nights in 2021 Jan (Figure \ref{fig:soar_phot}). As seen in the earlier OGLE $I$ data, the source is strongly variable, with an even higher amplitude of $\sim 1.8$ mag.

\begin{figure*}[th!]
\begin{center}
\includegraphics[width=6.3in]{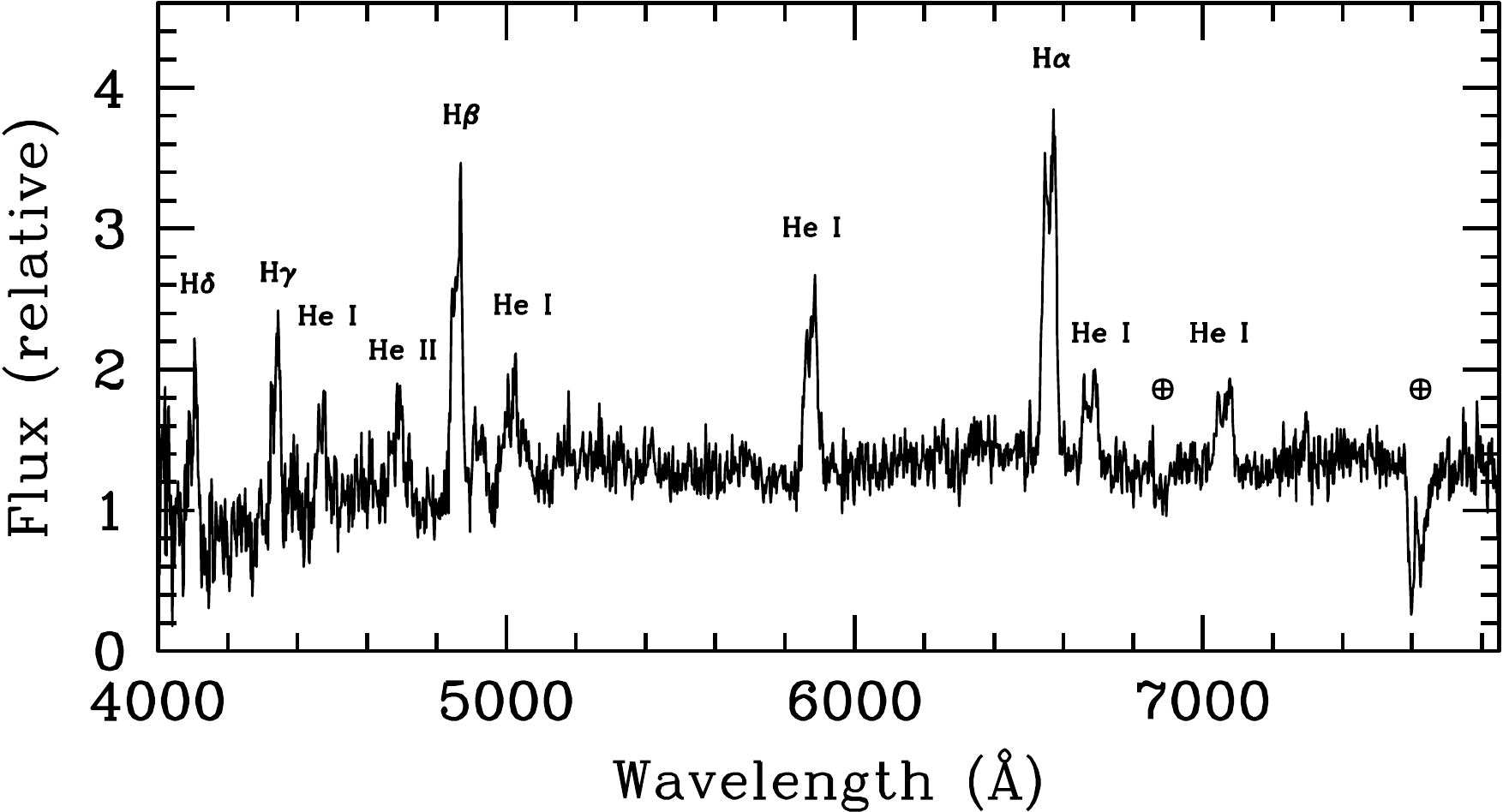}
\caption{A single 25-min SOAR/Goodman spectrum of J0540B from 2020 Dec 20, with a relative flux calibration applied. The double-peaked H and He emission lines characteristic of low-mass X-ray binaries are evident.}
\label{fig:specs}
\end{center}
\end{figure*}

However, unlike the case with the OGLE data, there is strong evidence for a periodic signal in the SOAR photometry via a Lomb-Scargle periodigram (with a nominal significance of $11\sigma$). The full (two night) dataset has an inferred period of $P = 0.1106\pm0.0026$ d. This period is also significantly detected in periodigrams of the two individual nights, with periods at $P = 0.109$ d and 0.114 d, respectively. In the light curves plotted in Figure \ref{fig:soar_phot}, this signal is readily apparent as the high-amplitude, low frequency variation on both nights. 

Figure \ref{fig:soar_phot} also shows high-frequency variations that recur quasi-periodically on timescales of $\sim 40$--45 min, especially on the second night. This timescale is similar to that observed for the X-ray flares in the \emph{XMM} data (Section 3.1) and hence it is reasonable to conclude that the two phenomena are linked.

\subsection{UV Photometry: XMM}

J0540B is not detected or only marginally detected in the individual ultraviolet \emph{XMM} OM images, but is cleanly detected in the stack of images, with $uvm2 = 22.40\pm0.13$ mag (on the AB system). This value corresponds to an extinction-corrected value of $uvm2_0 = 21.74$ mag, keeping in mind that this filter is near the UV dust bump at 2175 \AA\ and hence the reddening correction could have a substantial systematic uncertainty. The median $G$ mag from our SOAR photometry, transformed to the AB system and corrected for reddening, is $G_0 = 20.24$ mag. These were obtained $\sim 4$--10 d before the \emph{XMM} data, so are not simultaneous, but should generally represent the mean state of the system around the same time. The resulting color is $(uvm2-G)_{0} \sim 1.5$, which is equivalent to a power-law flux distribution of $F_{\nu} \propto \nu^{-1.5}$. While this exact exponent should not be taken too literally because of the variability of J0540B and the uncertain reddening correction in the UV, it is nearly identical to the slope measured for the UV and blue fluxes of the accretion disk in the candidate tMSP 3FGL J0427.9--6704 \citep{Strader16}, and is generally consistent with the spectral energy distribution expected for an irradiated disk at these frequencies \citep{Frank02}.

\subsection{Optical Spectroscopy}

All the optical spectra show the same characteristics: a flat continuum (in $F_{\lambda}$), double-peaked or broad H and He emission lines, and a lack of non-telluric absorption lines. A sample SOAR/Goodman spectrum is shown in Figure \ref{fig:specs}.

In more detail, the Balmer series is present in all spectra (extending down to H$\epsilon$ in the bluer SOAR spectra), with Paschen lines also visible in the Gemini spectra. The mean full width at half maximum (FWHM) of H$\alpha$ = $1909\pm29$ km s$^{-1}$, with an equivalent width of $61.5\pm2.9$ \AA. The mean separation of the two peaks is around 960 km s$^{-1}$. All of the typical strong \ion{He}{1} lines observed in the optical spectra of low-mass X-ray binaries are also present, with the strongest being the line near 5875 \AA, which has a mean FWHM around 2155 km s$^{-1}$. The \ion{He}{2} line at 4686 \AA\ is clear in all of the spectra with sufficient signal-to-noise and is very broad, with a mean FWHM of 4030 km s$^{-1}$. This progressive increase of FWHM  for lines with higher ionization potentials is as expected for an accretion disk.

There are slight variations in the mean H$\alpha$ wavelength among the different spectra, but these appear to be more consistent with changes in the relative flux of the blue and red components of the line rather than orbital variations---at least in this relatively modest collection of spectra. The average wavelength corresponds to a barycentric systemic velocity of the binary of about $+110$ 
km s$^{-1}$, which we take as a preliminary estimate of this value that could be superseded with future data.

The FWHM of the emission lines in this source are higher than typical for neutron star low-mass X-ray binaries, well into the range typically observed for stellar-mass black holes \citep{Casares15}.
For example, a series of SOAR/Goodman spectra taken of PSR J1023+0038 in its current disk state have a mean H$\alpha$ FWHM of $1525\pm27$ km s$^{-1}$, substantially lower than for J0540B, even though PSR J1023+0038 has a relatively short 4.8 hr period.  This points to some combination of a very short period, more massive primary, or more edge-on orbit for J0540B  than typical neutron star low-mass X-ray binaries, which we explore in Section 5.4.1.

\subsection{$\gamma$-ray Results}

The optical and X-ray evidence that a transition in J0540B could have occurred in 2013, together with the \emph{Fermi}-LAT 4FGL DR2 classification of 4FGL J0540.0--7552 as variable, motivated us to examine the 4FGL DR2 light curve of 4FGL J0540.0--7552 \citep{4FGLDR2,4FGL}. 4FGL and 4FGL DR2 quantify variability using a  {\tt Variability\_Index}, which is a test statistic that for 
non-variable sources in 4FGL DR2 is predicted to have $\chi^2$ distribution with 9 degrees of freedom. Hence, the observed 4FGL J0540.0--7552 {\tt Variability\_Index} of 23.65 would have a formal probability of occurring by chance of $p=0.005$. The corresponding probability for 4FGL J0540.0--7552 in the 8-year 4FGL catalog was $p=0.0007$. In both catalogs variability is considered probable if $p < 0.01.$

\begin{figure}[t]
\includegraphics[width=3.4in]{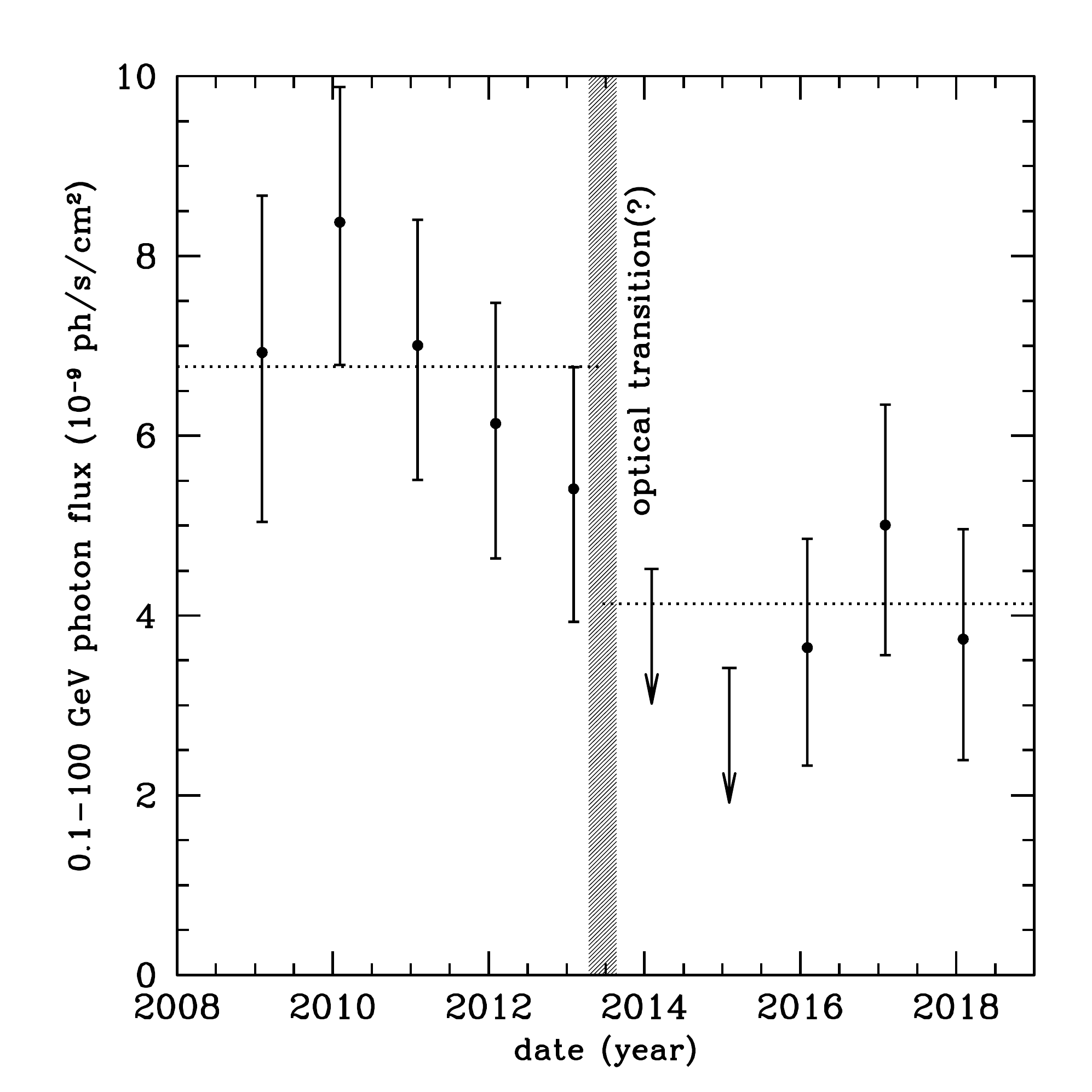}
\caption{Photon flux light curve for 4FGL J0540.0--7552 from 0.1--100 GeV, with 10 yearly bins from Aug 2008 to Aug 2018. The two upper limits are plotted at the the 95\% confidence level. The shaded region represents the seasonal gap for the optical OGLE coverage during which the optical transition appears to have occurred. The dashed lines are the mean photon fluxes for the pre-transition and post-transition time ranges, where the average of the three detections is conservatively plotted for the latter. }
\label{fig:ap_lc}
\end{figure}

The 4FGL DR2 light curve is shown in Figure \ref{fig:ap_lc}, where the 0.1--100 GeV photon flux is plotted in 1-yr bins. The bin edges occur in early August of each year. Since the optical transition appears to have occurred around the summer of 2013, and given the small number of \emph{Fermi} LAT photons likely associated with 4FGL J0540.0--7552 ($\lesssim 10$ per month), the five yearly bins before 2013 Aug can be reasonably associated with the pre-transition period, and the five yearly bins after 2013 Aug associated with the post-transition period.

4FGL J0540.0--7552 is well-detected ($\gtrsim 5\sigma$) in each of first five bins, through 2013 Aug. However, in the last five bins (from 2013 Aug to 2018 Aug), it is either not significantly detected 
or detected at a fainter flux (Figure \ref{fig:ap_lc}). If we conservatively take the post-transition photon flux as the mean value of the three detected bins, then the flux drops about 40\%, from 
$(6.8\pm0.5) \times 10^{-9}$ ph s$^{-1}$ cm$^{-2}$ to $(4.1\pm0.4) \times 10^{-9}$ ph s$^{-1}$ cm$^{-2}$. This change is formally significant at $4.1\sigma$. 

Given that the spectrum of 4FGL J0540.0--7552 peaks around 1 GeV, we checked to see whether any mismodeled sources that are bright around this energy could be affecting its light curve.
The only 4FGL DR2 catalog sources of comparable brightness within $5^{\circ}$ of 4FGL J0540.0--7552 are 4FGL J0558.8--7459 (separated by $1.5^{\circ}$; flux slightly lower than 4FGL J0540.0--7552 around 1 GeV) and 4FGL J0635.6--7518 (brighter at 1 GeV, but separated by $3.5^{\circ}$). Given their separations and fluxes, there is no evidence that mismodeling of these sources is contributing to the variability in 4FGL J0540.0--7552.

Overall, the \emph{Fermi}-LAT data are consistent with a sustained $\gamma$-ray flux change of 4FGL J0540.0--7552 having occurred at a time consistent with the optical flux change of J0540B.

\section{J0540A: An Unrelated RS CVn Binary}

J0540A is further from the center of the \emph{Fermi} error ellipse than J0540B, and, at least in the \emph{XMM} data, has a lower X-ray flux than J0540B. Nevertheless, it is within the 68\% \emph{Fermi} error ellipse, so for completeness, we discuss the properties of J0540A and why we believe it is unrelated to the $\gamma$-ray source 4FGL J0540.0--7552.

As mentioned in Section 2.3, J0540A is bright ($G=13.8$) and nearby ($925\pm10$ pc). It is also an optical variable, listed in the WISE variable star catalog as a Cepheid with a period of $2.489\pm0.011$ d \citep{Chen18}. Given an absolute mag of only $M_{G} \sim 3.8$, this classification is almost certainly incorrect, and as detailed below, a classification as an RS CVn star (a close active binary with an evolved component) is much more likely.

First, we checked the photometric period with data from the ASAS-SN survey \citep{Shappee14,Kochanek17}, using the over 4700 clean detections in $g$ obtained between 2017 Oct and 2021 Apr. The ASAS-SN mean magnitude is $g=14.7$. A Lomb-Scargle periodigram shows an extremely strong peak at $P = 2.4900\pm0.0002$ d, consistent with the WISE estimate. When the photometry is folded on this period, it shows a single peak with a slightly asymmetric shape, a mean peak to trough amplitude of about 0.22 mag, and substantial scatter at all phases. Examining the phased photometry as a function of time, there are clearly changes in the phase of the peak flux and the detailed shape of the light curve over time. Such changes are characteristic of rotational variability due to starspots on cool stars, which can vary in location and size over time \citep{Eaton79,Rodono95}.

As stated in Section 2.4, we also obtained SOAR spectroscopy of this star over several epochs in 2016 Sep to Dec. All the spectra are consistent with a cool star of mid-K spectral type, with no evidence for a second set of lines from a binary companion. A mid-K star with an absolute mag of $M_G \sim 3.8$ is evolved: a subgiant or a giant near the base of the red giant branch.

We derived barycentric radial velocities from these 14 spectra and fit a circular Keplerian model, finding a spectroscopic period consistent with the photometric one but with a lower precision. Hence we fixed the period to the photometric value, finding a satisfactory fit ($\chi^2$/d.o.f = 11.8/12) with a semi-amplitude of $K = 53.4\pm1.7$ km s$^{-1}$. The minimum mass of the unseen companion 
is $f(M) = PK^3/2\pi G = 0.039\pm0.003 M_{\odot}$. This is the mass function that would be expected for a wide range of reasonable parameters, e.g., an evolved primary of $\sim 1 M_{\odot}$ and a hidden, much fainter secondary of $\sim 0.5 M_{\odot}$ at the median inclination of $i \sim 60^{\circ}$. The velocities would not be consistent with a neutron star companion to the visible evolved star unless the system were nearly perfectly face-on. 

While dominated by narrow absorption lines, the optical spectra also all show H$\alpha$ emission which is slightly broadened, with a mean FWHM around $\sim 160$ km s$^{-1}$, but never double-peaked. The mean equivalent width of the H$\alpha$ emission is around $\sim 1.5$ \AA, typical for RS CVn stars \citep{Montes95}.

We found that the \emph{XMM} X-ray spectrum of J0540A was poorly fit by an absorbed power-law or disk blackbody. However, a soft absorbed power-law ($\Gamma = 2.7\pm0.4$) with two additional low-temperature thermal plasma ({\tt APEC}) components was a reasonable representation of the X-ray spectrum ($\chi^2$/d.o.f. = 58.0/61), though we do not claim this is necessarily a best-fit model. The inferred 0.3--8 keV X-ray luminosity from this model is $L_X = (1.5\pm0.5) \times 10^{31}$ erg s$^{-1}$. This model was preferred over one solely consisting of three {\tt APEC} components, which was an acceptable but less good fit ($\chi^2$/d.o.f. = 68.8/61).

There is no change in the count rate of the system during the \emph{XMM} observation, and the \emph{XMM} count rate is consistent with that predicted from the earlier \emph{Swift}/XRT observations (Section 2.2.1), suggesting that the X-ray data are representative of quiescence for this binary, rather than a flare. The X-ray luminosity and spectrum are consistent with that observed for RS CVn stars outside of flares \citep{Drake92,Pandey12}. 
 
Overall, the data are most consistent with the interpretation that J0540A is a binary with a spotted, cool, evolved primary and a lower-mass unseen secondary. The 2.49 d rotational period of the primary inferred from photometry is consistent with the orbital period inferred from optical spectroscopy, indicating that it is tidally locked to the secondary or nearly so. The binary shows evidence for  optical and X-ray emission of chromospheric and coronal origin. In short, all observed properties agree with a classification as an RS CVn-type active binary. Such binaries have never been observed to produce GeV $\gamma$-ray emission, and hence we conclude that J0540A is unrelated to the \emph{Fermi}-LAT $\gamma$-ray source.

\section{Discussion}

\subsection{Classification of 4FGL J0540.0--7552 a Candidate Transitional Millisecond Pulsar}

Returning to J0540B, the combination of the X-ray spectrum (a hard power-law with $\Gamma = 1.8$), extreme X-ray and optical variability, and the persistent presence of an optical accretion disk already peg J0540B as a low-mass X-ray binary with unusual properties. This source also sits within the 68\% error ellipse of a long-unassociated \emph{Fermi}-LAT source, and is the brightest X-ray source within the 68\% (or 95\%) error ellipse of the GeV $\gamma$-ray source. There is only one other significant X-ray source within the  \emph{Fermi}-LAT error ellipse, and the detailed study of this source (J0540A) in Section 4 shows that it is unrelated to the $\gamma$-ray source. Hence it is likely that the X-ray/optical source J0540B is indeed associated with 4FGL J0540.0--7552.

The only known source type that shares this combination of optical, X-ray, and $\gamma$-ray properties is the class of tMSPs while in the sub-luminous disk state. While only two field (and one globular cluster) sources are members of this class with confirmed state transitions between the pulsar and disk states, there are four additional published field sources that are strong tMSP candidates in the disk state. Unlike these other candidates, 4FGL J0540.0--7552 has suggestive (but not yet conclusive) evidence for having undergone a transition to its current 
sub-luminous disk state, which we discuss in more detail in Section 5.3.

\citet{Miller20} discussed using the ratio of X-ray flux ($F_X$; 0.5--10 keV) to $\gamma$-ray flux ($F_{\gamma}$; 0.1--100 GeV) to separate black widows and redbacks in the pulsar state from tMSPs in the sub-luminous disk state, especially relevant for sources with unknown distances. For 4FGL J0540.0--7552, we calculate a post-transition 0.1--100 GeV flux of $(2.9\pm0.5) \times 10^{-12}$ erg s$^{-1}$ cm$^{-2}$ by scaling the 4FGL DR2 integrated energy flux down using the post-transition photon flux. This gives $F_X/F_{\gamma} = 0.11\pm0.02$. This ratio is lower than that previously measured for confirmed or candidate tMSPs in the sub-luminous disk state, which ranged from $F_X/F_{\gamma} = 0.26$--0.43. While the ratio for 4FGL J0540.0--7552 is still much higher than typical for redbacks in the pulsar state (which have a median $F_X/F_{\gamma} = 0.01$), there may be a broader range of $F_X/F_{\gamma}$ values in the sub-luminous disk state than previously observed.

\subsection{Flare-Mode Transitional Millisecond Pulsars}

In its current sub-luminous disk state, the most intriguing aspect of its phenomenology is that 4FGL J0540.0--7552 was observed to spend essentially the entire \emph{XMM} observation in a flaring mode. One other candidate tMSP, 3FGL J0427.9--6704, is also dominated by flares. In detail, the properties of the flares in these two systems are somewhat different: 3FGL J0427.9--6704 has more frequent but shorter duration flares, with about 43\% of time spent in a flaring state, while 4FGL J0540.0--7552 has less frequent, longer duration flares, and a larger fraction of time ($\gtrsim 67$\%) spent in an active flare. Other candidate tMSPs, most especially PSR J1023+0038 \citep{Tendulkar14,Papitto18}, do show evidence for luminous flares and indeed periods of enhanced flaring, but none are dominated by flaring to the same consistent extent as 3FGL J0427.9--6704 or 4FGL J0540.0--7552. The OGLE photometry of 4FGL J0540.0--7552 from 2014/2015 (Figure \ref{fig:ogle}), showing extreme optical variability like that observed in 2021 around the time of the \emph{XMM} X-ray observations, suggests that 4FGL J0540.0--7552 exhibits this flaring mode consistently.

While we have no simultaneous X-ray and optical data for 4FGL J0540.0--7552 (the system was too faint to monitor at high cadence with the \emph{XMM} OM), there were optical flares in the SOAR photometry of J0540B that occurred on similar quasi-periodic $\sim 40$--45 min timescales as the X-ray flares in the \emph{XMM} data. This is consistent with a model in which the bulk of the optical flares represent X-ray flares reprocessed by the accretion disk, as was hypothesized for 3FGL J0427.9--6704 \citep{Li20}, though other models are possible.

As mentioned in \citet{Li20}, another published X-ray binary---source ``B" in the globular cluster NGC 6652---has $L_X = 1-2\times10^{34}$ erg s$^{-1}$, and exhibits repeated X-ray and optical flares, with light curves broadly similar to those of 4FGL J0540.0--7552 and 3FGL J0427.9--6704 \citep{Coomber11,Stacey12,Engel12}. This source also shows broad Balmer emission in optical spectroscopy as well as radio continuum emission \citep{Paduano21}. 

Based on the existing evidence of similar phenomenology, it seems reasonable to tentatively classify these three sources as ``flare-mode" candidate tMSPs, keeping in mind that the relationship between these sources and the confirmed tMSPs is still uncertain. It is possible that some of the simpler models initially used to explain tMSPs, such as a cyclic instability in the inner radius of the accretion disk \citep{DAngelo10}, could be relevant for understanding flare-mode tMSPs.

\subsection{Evidence for a Transition}

Only three sources have truly ``earned" the tMSP moniker through the observation of distinct pulsar and disk states in the same sources. The remainder of the published candidate tMSPs (3FGL J1544.6--1125,  3FGL J0427.9--6704, CXOU J110926.4--650224, 4FGL J0407.7--5702) match a subset of the characteristics of tMSPs in the sub-luminous disk state: optical evidence for an accretion disk; unusual X-ray variability (sometimes including distinct low/high X-ray modes); a hard power-law spectrum with $\Gamma \sim 1.5$--1.8; radio continuum emission; GeV $\gamma$-ray emission. None have shown evidence for a transition.

4FGL J0540.0--7552 is different: uniquely among the candidate tMSPs, it shows evidence for a transition in three distinct wavebands: the optical, X-ray and $\gamma$-ray. In the optical, J0540B brightened by 0.6 mag (a factor of 1.7) in $I$ photometry from OGLE in 2013. In the X-rays, J0540B brightened by a factor of $\sim 3$--7 sometime between 2012 May and 2021 Jan, due either
to an overall change in its X-ray brightness or to a dramatic increase in the occurrence of X-ray flares. An older \emph{ROSAT} non-detection also suggests its X-ray properties in the early 1990s were more consistent with those in 2012 than in 2021. Finally, there is evidence that 4FGL J0540.0--7552 faded by $\sim 40$\% in the 0.1--100 GeV band around the time of the 2013 OGLE optical change.

In detail, there are some differences between the pre/post transition properties of 4FGL J0540.0--7552 compared to the two well-studied field tMSPs PSR J1023+0038 and XSS J12270--4859. In the latter cases the X-ray luminosity shows a larger difference in the sub-luminous disk state compared to the pulsar state (a factor of $\sim 15$--30; \citealt{Bogdanov11,deMartino20}) than in J0540B. On the other hand, depending on their intrabinary shock properties, redbacks vary enormously in their pulsar state X-ray luminosity (from $\sim 10^{30}$--$10^{33}$ erg s$^{-1}$, \citealt{Roberts18,Swihart18,Strader19}), so a wide range in pre/post transition X-ray luminosities might also be expected. 

Another difference is that the known tMSPs are a factor of $\sim 3$--6 brighter in 0.1--100 GeV $\gamma$-rays in the sub-luminous disk state compared to the pulsar state \citep{Stappers14,Johnson15}, but 4FGL J0540.0--7552 is \emph{fainter} in the GeV $\gamma$-rays after the possible 2013 transition. A number of results about the sub-luminous disk state are counterintuitive, such as the finding that PSR J1023+0038 is spinning down more quickly in this state than in the pulsar state \citep{Jaodand16}. Given this, and that the origin of the $\gamma$-rays in the sub-luminous disk state is unknown, it seems plausible that even the direction of the change in $\gamma$-ray flux post-transition might vary among tMSPs. For example, if the flare-mode tMSPs are indeed accreting onto the neutron star at least some of the time, then this could lead to a reduced rather than enhanced spin-down rate and hence a lower $\gamma$-ray luminosity.

We also see no clear orbital period in the pre-transition OGLE $I$ photometry of J0540B, as might be expected if the system was in the pulsar state at this time. There are a number of reasons that periodic variability might be difficult to see even if present, including the small number of measurements, their large photometric uncertainties, and the incompleteness of OGLE photometry at these faint magnitudes.

It is also possible that a simple bimodal picture of a pulsar or disk state may not appropriately capture the phenomenology of tMSPs---perhaps there are multiple disk states, with transitions among these disk states sometimes occurring. ``Flare-mode" tMSPs like 4FGL J0540.0--7552 and 3FGL J0427.9--6704 could speculatively represent a different disk state than observed in typical tMSPs.

\subsection{Binary Properties}

\subsubsection{Orbital Period and Donor}

We identified a consistent periodic signal at $\sim 2.7$ hr in our SOAR photometry of J0540B on two separate nights six days apart. This signal could in principle represent either the orbital period, half the orbital period (if due to ellipsoidal variations), or could be spurious. If it were interpreted as the orbital period, it would be the shortest of any tMSP (or candidate) and indeed the shortest of any redback \citep{Strader19}. Nonetheless, there are black widows with similar orbital periods (e.g., \citealt{Bhattacharyya13}). While the known tMSPs are all redbacks, it is not obvious \emph{a priori} that a subset of black widows might not also undergo state transitions.

Another piece of evidence that a short period might be viable is the high FWHM of the emission lines compared to most neutron star low-mass X-ray binaries (Section 3.6). This points to a shorter period, more massive primary, or a more edge-on orbit than typical. Given the lack of obvious X-ray or optical eclipses, the inclination cannot be \emph{too} edge-on, likely $\lesssim 75^{\circ}$ \citep{Chanan76}. By contrast, there is substantial leverage in the period: if the emission lines are produced at a characteristic fraction of the primary Roche lobe radius, then one expects the emission line FWHM $\propto P^{-1/3}$. For example, a binary with a 2.7 hr orbital period would have a predicted FWHM about 20\% higher than that of PSR J1023+0038, 4.8 hr. This could explain most of the difference in the observed H$\alpha$ FWHM between these two systems (1909 km s$^{-1}$ for J0540B vs. 1525 km s$^{-1}$ for PSR J1023+0038) if their inclinations and neutron star masses are broadly similar.

If the secondary is of low mass and in a short period orbit, it might contribute to the continuum flux at superior conjunction when its warm heated face is visible, contributing to a photometric orbital signal. If sufficiently warm, it might show weak or no metal lines, with any hydrogen lines hidden by the stronger disk lines.

A 5.3 hr orbit (twice the periodic signal) would be a typical redback orbital period. However, given that no absorption lines are seen in the optical spectra, it is hard to see how ellipsoidal variations are tenable as the primary source of optical variability, since they would require a dominant contribution to the total flux from the companion star.

For some candidate tMSPs in the sub-luminous disk state, such as 3FGL J1544.6--1125, it was not possible to measure an orbital period even with extensive optical photometry \citep{BH15}; it required the detection of absorption lines from the secondary in optical spectroscopy \citep{Britt17}. Hence it is possible that our candidate period for 4FGL J0540.0--7552 is not genuine. Given that it would represent a new area of parameter space---either the shortest period redback, or a black widow with tMSP-like behavior---follow-up work to better constrain the period would be valuable.

\subsubsection{Distance and Velocity}

J0540B does not have a significant \emph{Gaia} parallax. A rough distance estimates can be made by comparing its unabsorbed 0.5--10 keV X-ray flux to the X-ray luminosities of the four known or candidate tMSPs with well-constrained distances, which lie in the range 1.6--$7.7 \times 10^{33}$ erg s$^{-1}$ \citep{Strader19,Li20}. This suggests a distance in the range $\sim 6.5$--14.2 kpc. A similar comparison of 0.1--100 GeV fluxes gives a partially overlapping distance range of 4.2--8.3 kpc. Another method, first performed for 4FGL J0407.7--5702 \citep{Miller20}, notes that the absolute \emph{Gaia} $BP$ mag of the  tMSP sample has a relatively narrow range ($\sim 5.6$--6.0).  Comparing this to the mean $BP_0$ for J0540B ($20.20\pm0.11$ mag) gives an estimated  optical distance in the range $\sim 6.6$--8.7 kpc. These distance estimates roughly overlap in the range $\sim 6.5$--8 kpc. While the true distance to J0540B is as yet unknown, a nearby distance of $\lesssim 4$--5 kpc seems unlikely.

At its large proper motion of $13.5\pm0.6$ mas yr$^{-1}$ and given its probable distance $\gtrsim 5$ kpc, J0540B seems likely to have a transverse velocity $\gtrsim 300$ km s$^{-1}$, which would be among the largest for any spider MSP binary \citep{Jennings18,Strader19}. The high space velocities of binary MSPs have been extensively studied (e.g., \citealt{Hobbs05,Desvignes16,Lynch18}), and \emph{Gaia} proper motions are likely to be among the tools useful for identifying new candidate spider MSPs (see also \citealt{Antoniadis21}).

\subsection{Future Work}

One wavelength regime with minimal coverage in this paper is the radio, with only shallow archival limits available for 4FGL J0540.0--7552. The majority of the known or candidate tMSPs in the sub-luminous disk state have been detected as radio continuum sources \citep{Hill11,Deller15,Jaodand19,Li20}. Of these, 3FGL J0427.9--6704 shows the most luminous radio continuum emission, and at all times sits close to the radio/X-ray correlation for black holes \citep{Li20}, which tend to be more radio-loud than accreting neutron stars (e.g., \citealt{Gallo18}). Furthermore, despite the optical, X-ray, and $\gamma$-ray eclipses observed for 3FGL J0427.9--6704 \citep{Strader16,Li20,Kennedy20}, the radio emission is not eclipsed, suggesting it is spatially extended \citep{Li20}. This steady, extended, flat-spectrum radio continuum emission has properties consistent with a jet. This is in dramatic contrast to the radio emission in PSR J1023+0038, which turns on and off on timescales as short as seconds and hence must be produced close to the neutron star rather than in a steady jet \citep{Bogdanov18}. This hints at the possibility that flare-mode tMSP candidates could have different radio continuum properties than typical tMSPs, and provides motivation for a deep radio continuum study of 4FGL J0540.0--7552, which could have a 5 GHz flux density as high as 0.1--0.2 mJy, depending on its distance.

More X-ray and optical data would also be useful: new X-ray observations would allow one to determine whether the system is always in the flaring mode, and new high-cadence optical photometry could help assess if the candidate orbital period discussed in this paper is real or spurious. Finally, as for all candidate tMSPs, multiwavelength surveillance of 4FGL J0540.0--7552 to search for evidence of a future transition is essential.

\begin{acknowledgments}

We thank an anonymous referee for comments that helped improve the paper.

JS acknowledges support from NSF grant AST-1714825 and the Packard Foundation. This research was performed while SJS held a NRC Research Associateship award at the Naval Research Laboratory. Work at the Naval Research Laboratory is supported by NASA DPR S-15633-Y. 

We thank the XMM-Newton Project Scientist, Observing Time Allocation Committee, and scheduling team for granting and efficiently scheduling discretionary time observations. This work was 
based on observations obtained with XMM-Newton, an ESA science mission with instruments and contributions directly funded by ESA Member States and NASA.

This work made use of data supplied by the UK Swift Science Data Centre at the University of Leicester.

Based on observations obtained at the Southern Astrophysical Research (SOAR) telescope, which is a joint project of the Minist\'{e}rio da Ci\^{e}ncia, Tecnologia e Inova\c{c}\~{o}es (MCTI/LNA) do Brasil, the US National Science Foundation's NOIRLab, the University of North Carolina at Chapel Hill (UNC), and Michigan State University (MSU).

Based on observations obtained at the international Gemini Observatory, a program of NSF's NOIRLab, which is managed by the Association of Universities for Research in Astronomy (AURA) under a cooperative agreement with the National Science Foundation on behalf of the Gemini Observatory partnership: the National Science Foundation (United States), National Research Council (Canada), Agencia Nacional de Investigaci\'{o}n y Desarrollo (Chile), Ministerio de Ciencia, Tecnolog\'{i}a e Innovaci\'{o}n (Argentina), Minist\'{e}rio da Ci\^{e}ncia, Tecnologia, Inova\c{c}\~{o}es e Comunica\c{c}\~{o}es (Brazil), and Korea Astronomy and Space Science Institute (Republic of Korea).

This research has made use of data and software provided by the High Energy Astrophysics Science Archive Research Center (HEASARC), which is a service of the Astrophysics Science Division
at NASA/GSFC and the High Energy Astrophysics Division of the Smithsonian Astrophysical Observatory.

The OGLE project has received funding from the National Science Centre, Poland, grant MAESTRO 2014/14/A/ST9/00121 to AU.

\end{acknowledgments}

\software{\emph{Swift}/XRT product tools \citep{Evans20}, SAS (v18.0.0; \citealt{2004ASPC..314..759G}), IRAF \citep{Tody86}, {\it lcmath} \citep{1995ASPC...77..367B}, XSPEC (v12.10.1; \citealt{1996ASPC..101...17A}), {\it lomb} \citep{Ruf99}, {\tt R} \citep{Rmanual}.}

\bibliographystyle{aasjournal}
\bibliography{j0540_arxiv}

\begin{thebibliography}{}
\expandafter\ifx\csname natexlab\endcsname\relax\def\natexlab#1{#1}\fi
\providecommand{\url}[1]{\href{#1}{#1}}
\providecommand{\dodoi}[1]{doi:~\href{http://doi.org/#1}{\nolinkurl{#1}}}
\providecommand{\doeprint}[1]{\href{http://ascl.net/#1}{\nolinkurl{http://ascl.net/#1}}}
\providecommand{\doarXiv}[1]{\href{https://arxiv.org/abs/#1}{\nolinkurl{https://arxiv.org/abs/#1}}}

\bibitem[{{Abdo} {et~al.}(2013){Abdo}, {Ajello}, {Allafort}, {Baldini},
  {Ballet}, {Barbiellini}, {Baring}, {Bastieri}, {Belfiore}, {Bellazzini}, \&
  et~al.}]{Abdo13}
{Abdo}, A.~A., {Ajello}, M., {Allafort}, A., {et~al.} 2013, \apjs, 208, 17,
  \dodoi{10.1088/0067-0049/208/2/17}

\bibitem[{{Abdollahi} {et~al.}(2020){Abdollahi}, {Acero}, {Ackermann},
  {Ajello}, {Atwood}, {Axelsson}, {Baldini}, {Ballet}, {Barbiellini},
  {Bastieri}, {Becerra Gonzalez}, {Bellazzini}, {Berretta}, {Bissaldi},
  {Blandford}, {Bloom}, {Bonino}, {Bottacini}, {Brandt}, {Bregeon}, {Bruel},
  {Buehler}, {Burnett}, {Buson}, {Cameron}, {Caputo}, {Caraveo}, {Casandjian},
  {Castro}, {Cavazzuti}, {Charles}, {Chaty}, {Chen}, {Cheung}, {Chiaro},
  {Ciprini}, {Cohen-Tanugi}, {Cominsky}, {Coronado-Bl{\'a}zquez}, {Costantin},
  {Cuoco}, {Cutini}, {D'Ammando}, {DeKlotz}, {de la Torre Luque}, {de Palma},
  {Desai}, {Digel}, {Di Lalla}, {Di Mauro}, {Di Venere}, {Dom{\'\i}nguez},
  {Dumora}, {Fana Dirirsa}, {Fegan}, {Ferrara}, {Franckowiak}, {Fukazawa},
  {Funk}, {Fusco}, {Gargano}, {Gasparrini}, {Giglietto}, {Giommi}, {Giordano},
  {Giroletti}, {Glanzman}, {Green}, {Grenier}, {Griffin}, {Grondin}, {Grove},
  {Guiriec}, {Harding}, {Hayashi}, {Hays}, {Hewitt}, {Horan},
  {J{\'o}hannesson}, {Johnson}, {Kamae}, {Kerr}, {Kocevski}, {Kovac'evic'},
  {Kuss}, {Landriu}, {Larsson}, {Latronico}, {Lemoine-Goumard}, {Li},
  {Liodakis}, {Longo}, {Loparco}, {Lott}, {Lovellette}, {Lubrano}, {Madejski},
  {Maldera}, {Malyshev}, {Manfreda}, {Marchesini}, {Marcotulli},
  {Mart{\'\i}-Devesa}, {Martin}, {Massaro}, {Mazziotta}, {McEnery}, {Mereu},
  {Meyer}, {Michelson}, {Mirabal}, {Mizuno}, {Monzani}, {Morselli},
  {Moskalenko}, {Negro}, {Nuss}, {Ojha}, {Omodei}, {Orienti}, {Orlando},
  {Ormes}, {Palatiello}, {Paliya}, {Paneque}, {Pei}, {Pe{\~n}a-Herazo},
  {Perkins}, {Persic}, {Pesce-Rollins}, {Petrosian}, {Petrov}, {Piron}, {Poon},
  {Porter}, {Principe}, {Rain{\`o}}, {Rando}, {Razzano}, {Razzaque}, {Reimer},
  {Reimer}, {Remy}, {Reposeur}, {Romani}, {Saz Parkinson}, {Schinzel},
  {Serini}, {Sgr{\`o}}, {Siskind}, {Smith}, {Spandre}, {Spinelli}, {Strong},
  {Suson}, {Tajima}, {Takahashi}, {Tak}, {Thayer}, {Thompson}, {Tibaldo},
  {Torres}, {Torresi}, {Valverde}, {Van Klaveren}, {van Zyl}, {Wood},
  {Yassine}, \& {Zaharijas}}]{4FGL}
{Abdollahi}, S., {Acero}, F., {Ackermann}, M., {et~al.} 2020, \apjs, 247, 33,
  \dodoi{10.3847/1538-4365/ab6bcb}

\bibitem[{{Alpar} {et~al.}(1982){Alpar}, {Cheng}, {Ruderman}, \&
  {Shaham}}]{Alpar82}
{Alpar}, M.~A., {Cheng}, A.~F., {Ruderman}, M.~A., \& {Shaham}, J. 1982, \nat,
  300, 728, \dodoi{10.1038/300728a0}

\bibitem[{{Andrew} {et~al.}(2021){Andrew}, {Swihart}, \& {Strader}}]{Andrew21}
{Andrew}, S., {Swihart}, S.~J., \& {Strader}, J. 2021, \apj, 908, 180,
  \dodoi{10.3847/1538-4357/abd257}

\bibitem[{{Antoniadis}(2021)}]{Antoniadis21}
{Antoniadis}, J. 2021, \mnras, 501, 1116, \dodoi{10.1093/mnras/staa3595}

\bibitem[{{Archibald} {et~al.}(2009){Archibald}, {Stairs}, {Ransom}, {Kaspi},
  {Kondratiev}, {Lorimer}, {McLaughlin}, {Boyles}, {Hessels}, {Lynch}, {van
  Leeuwen}, {Roberts}, {Jenet}, {Champion}, {Rosen}, {Barlow}, {Dunlap}, \&
  {Remillard}}]{Archibald09}
{Archibald}, A.~M., {Stairs}, I.~H., {Ransom}, S.~M., {et~al.} 2009, Science,
  324, 1411, \dodoi{10.1126/science.1172740}

\bibitem[{{Arnaud}(1996)}]{1996ASPC..101...17A}
{Arnaud}, K.~A. 1996, in Astronomical Society of the Pacific Conference Series,
  Vol. 101, Astronomical Data Analysis Software and Systems V, ed. G.~H.
  {Jacoby} \& J.~{Barnes}, 17

\bibitem[{{Bahramian} {et~al.}(2015){Bahramian}, {Heinke}, {Degenaar},
  {Chomiuk}, {Wijnands}, {Strader}, {Ho}, \& {Pooley}}]{Bahramian15}
{Bahramian}, A., {Heinke}, C.~O., {Degenaar}, N., {et~al.} 2015, \mnras, 452,
  3475, \dodoi{10.1093/mnras/stv1585}

\bibitem[{{Bahramian} {et~al.}(2018){Bahramian}, {Strader}, {Chomiuk},
  {Heinke}, {Miller-Jones}, {Degenaar}, {Tetarenko}, {Tudor}, {Tremou},
  {Shishkovsky}, {Wijnands}, {Maccarone}, {Sivakoff}, \&
  {Ransom}}]{Bahramian18}
{Bahramian}, A., {Strader}, J., {Chomiuk}, L., {et~al.} 2018, \apj, 864, 28,
  \dodoi{10.3847/1538-4357/aad68b}

\bibitem[{{Bahramian} {et~al.}(2020){Bahramian}, {Strader}, {Miller-Jones},
  {Chomiuk}, {Heinke}, {Maccarone}, {Pooley}, {Shishkovsky}, {Tudor}, {Zhao},
  {Li}, {Sivakoff}, {Tremou}, \& {Buchner}}]{Bahramian20}
{Bahramian}, A., {Strader}, J., {Miller-Jones}, J. C.~A., {et~al.} 2020, \apj,
  901, 57, \dodoi{10.3847/1538-4357/aba51d}

\bibitem[{{Bailer-Jones} {et~al.}(2021){Bailer-Jones}, {Rybizki}, {Fouesneau},
  {Demleitner}, \& {Andrae}}]{BJ21}
{Bailer-Jones}, C.~A.~L., {Rybizki}, J., {Fouesneau}, M., {Demleitner}, M., \&
  {Andrae}, R. 2021, \aj, 161, 147, \dodoi{10.3847/1538-3881/abd806}

\bibitem[{{Ballet} {et~al.}(2020){Ballet}, {Burnett}, {Digel}, \&
  {Lott}}]{4FGLDR2}
{Ballet}, J., {Burnett}, T.~H., {Digel}, S.~W., \& {Lott}, B. 2020, arXiv
  e-prints, arXiv:2005.11208.
\newblock \doarXiv{2005.11208}

\bibitem[{{Bassa} {et~al.}(2014){Bassa}, {Patruno}, {Hessels}, {Keane},
  {Monard}, {Mahony}, {Bogdanov}, {Corbel}, {Edwards}, {Archibald}, {Janssen},
  {Stappers}, \& {Tendulkar}}]{Bassa14}
{Bassa}, C.~G., {Patruno}, A., {Hessels}, J.~W.~T., {et~al.} 2014, \mnras, 441,
  1825, \dodoi{10.1093/mnras/stu708}

\bibitem[{{Bhattacharya} \& {van den Heuvel}(1991)}]{Bhattacharya91}
{Bhattacharya}, D., \& {van den Heuvel}, E.~P.~J. 1991, \physrep, 203, 1,
  \dodoi{10.1016/0370-1573(91)90064-S}

\bibitem[{{Bhattacharyya} {et~al.}(2013){Bhattacharyya}, {Roy}, {Ray}, {Gupta},
  {Bhattacharya}, {Romani}, {Ransom}, {Ferrara}, {Wolff}, {Camilo}, {Cognard},
  {Harding}, {den Hartog}, {Johnston}, {Keith}, {Kerr}, {Michelson}, {Saz
  Parkinson}, {Wood}, \& {Wood}}]{Bhattacharyya13}
{Bhattacharyya}, B., {Roy}, J., {Ray}, P.~S., {et~al.} 2013, \apjl, 773, L12,
  \dodoi{10.1088/2041-8205/773/1/L12}

\bibitem[{{Blackburn}(1995)}]{1995ASPC...77..367B}
{Blackburn}, J.~K. 1995, in Astronomical Society of the Pacific Conference
  Series, Vol.~77, Astronomical Data Analysis Software and Systems IV, ed.
  R.~A. {Shaw}, H.~E. {Payne}, \& J.~J.~E. {Hayes}, 367

\bibitem[{{Bogdanov} {et~al.}(2011){Bogdanov}, {Archibald}, {Hessels}, {Kaspi},
  {Lorimer}, {McLaughlin}, {Ransom}, \& {Stairs}}]{Bogdanov11}
{Bogdanov}, S., {Archibald}, A.~M., {Hessels}, J. W.~T., {et~al.} 2011, \apj,
  742, 97, \dodoi{10.1088/0004-637X/742/2/97}

\bibitem[{{Bogdanov} \& {Halpern}(2015)}]{BH15}
{Bogdanov}, S., \& {Halpern}, J.~P. 2015, \apjl, 803, L27,
  \dodoi{10.1088/2041-8205/803/2/L27}

\bibitem[{{Bogdanov} {et~al.}(2015){Bogdanov}, {Archibald}, {Bassa}, {Deller},
  {Halpern}, {Heald}, {Hessels}, {Janssen}, {Lyne}, {Mold{\'o}n}, {Paragi},
  {Patruno}, {Perera}, {Stappers}, {Tendulkar}, {D'Angelo}, \& {Wijnand
  s}}]{Bogdanov15}
{Bogdanov}, S., {Archibald}, A.~M., {Bassa}, C., {et~al.} 2015, \apj, 806, 148,
  \dodoi{10.1088/0004-637X/806/2/148}

\bibitem[{{Bogdanov} {et~al.}(2018){Bogdanov}, {Deller}, {Miller-Jones},
  {Archibald}, {Hessels}, {Jaodand}, {Patruno}, {Bassa}, \&
  {D'Angelo}}]{Bogdanov18}
{Bogdanov}, S., {Deller}, A.~T., {Miller-Jones}, J. C.~A., {et~al.} 2018, \apj,
  856, 54, \dodoi{10.3847/1538-4357/aaaeb9}

\bibitem[{{Boller} {et~al.}(2016){Boller}, {Freyberg}, {Tr{\"u}mper}, {Haberl},
  {Voges}, \& {Nandra}}]{ROSAT}
{Boller}, T., {Freyberg}, M.~J., {Tr{\"u}mper}, J., {et~al.} 2016, \aap, 588,
  A103, \dodoi{10.1051/0004-6361/201525648}

\bibitem[{{Britt} {et~al.}(2017){Britt}, {Strader}, {Chomiuk}, {Tremou},
  {Peacock}, {Halpern}, \& {Salinas}}]{Britt17}
{Britt}, C.~T., {Strader}, J., {Chomiuk}, L., {et~al.} 2017, \apj, 849, 21,
  \dodoi{10.3847/1538-4357/aa8e41}

\bibitem[{{Campana} {et~al.}(2019){Campana}, {Miraval Zanon}, {Coti Zelati},
  {Torres}, {Baglio}, \& {Papitto}}]{Campana19}
{Campana}, S., {Miraval Zanon}, A., {Coti Zelati}, F., {et~al.} 2019, \aap,
  629, L8, \dodoi{10.1051/0004-6361/201936312}

\bibitem[{{Casares}(2015)}]{Casares15}
{Casares}, J. 2015, \apj, 808, 80, \dodoi{10.1088/0004-637X/808/1/80}

\bibitem[{{Chanan} {et~al.}(1976){Chanan}, {Middleditch}, \&
  {Nelson}}]{Chanan76}
{Chanan}, G.~A., {Middleditch}, J., \& {Nelson}, J.~E. 1976, \apj, 208, 512,
  \dodoi{10.1086/154633}

\bibitem[{{Chen} {et~al.}(2018){Chen}, {Wang}, {Deng}, {de Grijs}, \&
  {Yang}}]{Chen18}
{Chen}, X., {Wang}, S., {Deng}, L., {de Grijs}, R., \& {Yang}, M. 2018, \apjs,
  237, 28, \dodoi{10.3847/1538-4365/aad32b}

\bibitem[{{Clemens} {et~al.}(2004){Clemens}, {Crain}, \&
  {Anderson}}]{Clemens04}
{Clemens}, J.~C., {Crain}, J.~A., \& {Anderson}, R. 2004, in Society of
  Photo-Optical Instrumentation Engineers (SPIE) Conference Series, Vol. 5492,
  Ground-based Instrumentation for Astronomy, ed. A.~F.~M. {Moorwood} \&
  M.~{Iye}, 331--340

\bibitem[{{Coomber} {et~al.}(2011){Coomber}, {Heinke}, {Cohn}, {Lugger}, \&
  {Grindlay}}]{Coomber11}
{Coomber}, G., {Heinke}, C.~O., {Cohn}, H.~N., {Lugger}, P.~M., \& {Grindlay},
  J.~E. 2011, \apj, 735, 95, \dodoi{10.1088/0004-637X/735/2/95}

\bibitem[{{Coti Zelati} {et~al.}(2019){Coti Zelati}, {Papitto}, {de Martino},
  {Buckley}, {Odendaal}, {Li}, {Russell}, {Torres}, {Mazzola}, {Bozzo},
  {Gromadzki}, {Campana}, {Rea}, {Ferrigno}, \& {Migliari}}]{CotiZelati19}
{Coti Zelati}, F., {Papitto}, A., {de Martino}, D., {et~al.} 2019, \aap, 622,
  A211, \dodoi{10.1051/0004-6361/201834835}

\bibitem[{{D'Angelo} \& {Spruit}(2010)}]{DAngelo10}
{D'Angelo}, C.~R., \& {Spruit}, H.~C. 2010, \mnras, 406, 1208,
  \dodoi{10.1111/j.1365-2966.2010.16749.x}

\bibitem[{{de Martino} {et~al.}(2020){de Martino}, {Papitto}, {Burgay},
  {Possenti}, {Coti Zelati}, {Rea}, {Torres}, \& {Belloni}}]{deMartino20}
{de Martino}, D., {Papitto}, A., {Burgay}, M., {et~al.} 2020, \mnras, 492,
  5607, \dodoi{10.1093/mnras/staa164}

\bibitem[{{de Martino} {et~al.}(2013){de Martino}, {Belloni}, {Falanga},
  {Papitto}, {Motta}, {Pellizzoni}, {Evangelista}, {Piano}, {Masetti},
  {Bonnet-Bidaud}, {Mouchet}, {Mukai}, \& {Possenti}}]{deMartino13}
{de Martino}, D., {Belloni}, T., {Falanga}, M., {et~al.} 2013, \aap, 550, A89,
  \dodoi{10.1051/0004-6361/201220393}

\bibitem[{{Deller} {et~al.}(2015){Deller}, {Moldon}, {Miller-Jones}, {Patruno},
  {Hessels}, {Archibald}, {Paragi}, {Heald}, \& {Vilchez}}]{Deller15}
{Deller}, A.~T., {Moldon}, J., {Miller-Jones}, J.~C.~A., {et~al.} 2015, \apj,
  809, 13, \dodoi{10.1088/0004-637X/809/1/13}

\bibitem[{{Desvignes} {et~al.}(2016){Desvignes}, {Caballero}, {Lentati},
  {Verbiest}, {Champion}, {Stappers}, {Janssen}, {Lazarus}, {Os{\l}owski},
  {Babak}, {Bassa}, {Brem}, {Burgay}, {Cognard}, {Gair}, {Graikou},
  {Guillemot}, {Hessels}, {Jessner}, {Jordan}, {Karuppusamy}, {Kramer},
  {Lassus}, {Lazaridis}, {Lee}, {Liu}, {Lyne}, {McKee}, {Mingarelli},
  {Perrodin}, {Petiteau}, {Possenti}, {Purver}, {Rosado}, {Sanidas}, {Sesana},
  {Shaifullah}, {Smits}, {Taylor}, {Theureau}, {Tiburzi}, {van Haasteren}, \&
  {Vecchio}}]{Desvignes16}
{Desvignes}, G., {Caballero}, R.~N., {Lentati}, L., {et~al.} 2016, \mnras, 458,
  3341, \dodoi{10.1093/mnras/stw483}

\bibitem[{{Drake} {et~al.}(1992){Drake}, {Simon}, \& {Linsky}}]{Drake92}
{Drake}, S.~A., {Simon}, T., \& {Linsky}, J.~L. 1992, \apjs, 82, 311,
  \dodoi{10.1086/191717}

\bibitem[{{Eastman} {et~al.}(2010){Eastman}, {Siverd}, \& {Gaudi}}]{Eastman10}
{Eastman}, J., {Siverd}, R., \& {Gaudi}, B.~S. 2010, \pasp, 122, 935,
  \dodoi{10.1086/655938}

\bibitem[{{Eaton} \& {Hall}(1979)}]{Eaton79}
{Eaton}, J.~A., \& {Hall}, D.~S. 1979, \apj, 227, 907, \dodoi{10.1086/156800}

\bibitem[{{Engel} {et~al.}(2012){Engel}, {Heinke}, {Sivakoff}, {Elshamouty}, \&
  {Edmonds}}]{Engel12}
{Engel}, M.~C., {Heinke}, C.~O., {Sivakoff}, G.~R., {Elshamouty}, K.~G., \&
  {Edmonds}, P.~D. 2012, \apj, 747, 119, \dodoi{10.1088/0004-637X/747/2/119}

\bibitem[{{Evans} {et~al.}(2020){Evans}, {Page}, {Osborne}, {Beardmore},
  {Willingale}, {Burrows}, {Kennea}, {Perri}, {Capalbi}, {Tagliaferri}, \&
  {Cenko}}]{Evans20}
{Evans}, P.~A., {Page}, K.~L., {Osborne}, J.~P., {et~al.} 2020, \apjs, 247, 54,
  \dodoi{10.3847/1538-4365/ab7db9}

\bibitem[{{Frank} {et~al.}(2002){Frank}, {King}, \& {Raine}}]{Frank02}
{Frank}, J., {King}, A., \& {Raine}, D.~J. 2002, {Accretion Power in
  Astrophysics: Third Edition} (Cambridge, UK: Cambridge University Press)

\bibitem[{{Gabriel} {et~al.}(2004){Gabriel}, {Denby}, {Fyfe}, {Hoar}, {Ibarra},
  {Ojero}, {Osborne}, {Saxton}, {Lammers}, \& {Vacanti}}]{2004ASPC..314..759G}
{Gabriel}, C., {Denby}, M., {Fyfe}, D.~J., {et~al.} 2004, in Astronomical
  Society of the Pacific Conference Series, Vol. 314, Astronomical Data
  Analysis Software and Systems (ADASS) XIII, ed. F.~{Ochsenbein}, M.~G.
  {Allen}, \& D.~{Egret}, 759

\bibitem[{{Gaia Collaboration} {et~al.}(2020){Gaia Collaboration}, {Brown},
  {Vallenari}, {Prusti}, {de Bruijne}, {Babusiaux}, \& {Biermann}}]{EDR3}
{Gaia Collaboration}, {Brown}, A.~G.~A., {Vallenari}, A., {et~al.} 2020, arXiv
  e-prints, arXiv:2012.01533.
\newblock \doarXiv{2012.01533}

\bibitem[{{Gallo} {et~al.}(2018){Gallo}, {Degenaar}, \& {van den
  Eijnden}}]{Gallo18}
{Gallo}, E., {Degenaar}, N., \& {van den Eijnden}, J. 2018, \mnras, 478, L132,
  \dodoi{10.1093/mnrasl/sly083}

\bibitem[{{G{\"u}ver} \& {{\"O}zel}(2009)}]{Guver09}
{G{\"u}ver}, T., \& {{\"O}zel}, F. 2009, \mnras, 400, 2050,
  \dodoi{10.1111/j.1365-2966.2009.15598.x}

\bibitem[{HEASARC(2014)}]{2014ascl.soft08004N}
HEASARC. 2014, {HEAsoft: Unified Release of FTOOLS and XANADU}.
\newblock \doeprint{1408.004}

\bibitem[{{Hill} {et~al.}(2011){Hill}, {Szostek}, {Corbel}, {Camilo}, {Corbet},
  {Dubois}, {Dubus}, {Edwards}, {Ferrara}, {Kerr}, {Koerding}, {Kozie{\l}}, \&
  {Stawarz}}]{Hill11}
{Hill}, A.~B., {Szostek}, A., {Corbel}, S., {et~al.} 2011, \mnras, 415, 235,
  \dodoi{10.1111/j.1365-2966.2011.18692.x}

\bibitem[{{Hobbs} {et~al.}(2005){Hobbs}, {Lorimer}, {Lyne}, \&
  {Kramer}}]{Hobbs05}
{Hobbs}, G., {Lorimer}, D.~R., {Lyne}, A.~G., \& {Kramer}, M. 2005, \mnras,
  360, 974, \dodoi{10.1111/j.1365-2966.2005.09087.x}

\bibitem[{{Hooper} \& {Linden}(2016)}]{Hooper16}
{Hooper}, D., \& {Linden}, T. 2016, \jcap, 2016, 018,
  \dodoi{10.1088/1475-7516/2016/08/018}

\bibitem[{{Jaodand} {et~al.}(2016){Jaodand}, {Archibald}, {Hessels},
  {Bogdanov}, {D'Angelo}, {Patruno}, {Bassa}, \& {Deller}}]{Jaodand16}
{Jaodand}, A., {Archibald}, A.~M., {Hessels}, J. W.~T., {et~al.} 2016, \apj,
  830, 122, \dodoi{10.3847/0004-637X/830/2/122}

\bibitem[{{Jaodand}(2019)}]{Jaodand19}
{Jaodand}, A.~D. 2019, PhD thesis, University of Amsterdam

\bibitem[{{Jennings} {et~al.}(2018){Jennings}, {Kaplan}, {Chatterjee},
  {Cordes}, \& {Deller}}]{Jennings18}
{Jennings}, R.~J., {Kaplan}, D.~L., {Chatterjee}, S., {Cordes}, J.~M., \&
  {Deller}, A.~T. 2018, \apj, 864, 26, \dodoi{10.3847/1538-4357/aad084}

\bibitem[{{Johnson} {et~al.}(2015){Johnson}, {Ray}, {Roy}, {Cheung}, {Harding},
  {Pletsch}, {Fort}, {Camilo}, {Deneva}, {Bhattacharyya}, {Stappers}, \&
  {Kerr}}]{Johnson15}
{Johnson}, T.~J., {Ray}, P.~S., {Roy}, J., {et~al.} 2015, \apj, 806, 91,
  \dodoi{10.1088/0004-637X/806/1/91}

\bibitem[{{Kennedy} {et~al.}(2020){Kennedy}, {Breton}, {Clark}, {Dhillon},
  {Kerr}, {Buckley}, {Potter}, {Mata S{\'a}nchez}, {Stringer}, \&
  {Marsh}}]{Kennedy20}
{Kennedy}, M.~R., {Breton}, R.~P., {Clark}, C.~J., {et~al.} 2020, \mnras, 494,
  3912, \dodoi{10.1093/mnras/staa912}

\bibitem[{{Kochanek} {et~al.}(2017){Kochanek}, {Shappee}, {Stanek}, {Holoien},
  {Thompson}, {Prieto}, {Dong}, {Shields}, {Will}, {Britt}, {Perzanowski}, \&
  {Pojma{\'n}ski}}]{Kochanek17}
{Kochanek}, C.~S., {Shappee}, B.~J., {Stanek}, K.~Z., {et~al.} 2017, \pasp,
  129, 104502, \dodoi{10.1088/1538-3873/aa80d9}

\bibitem[{{Li} {et~al.}(2020){Li}, {Strader}, {Miller-Jones}, {Heinke}, \&
  {Chomiuk}}]{Li20}
{Li}, K.-L., {Strader}, J., {Miller-Jones}, J. C.~A., {Heinke}, C.~O., \&
  {Chomiuk}, L. 2020, \apj, 895, 89, \dodoi{10.3847/1538-4357/ab8f28}

\bibitem[{{Linares} {et~al.}(2014){Linares}, {Bahramian}, {Heinke}, {Wijnand
  s}, {Patruno}, {Altamirano}, {Homan}, {Bogdanov}, \& {Pooley}}]{Linares14}
{Linares}, M., {Bahramian}, A., {Heinke}, C., {et~al.} 2014, \mnras, 438, 251,
  \dodoi{10.1093/mnras/stt2167}

\bibitem[{{Lindegren} {et~al.}(2020){Lindegren}, {Bastian}, {Biermann},
  {Bombrun}, {de Torres}, {Gerlach}, {Geyer}, {Hern{\'a}ndez}, {Hilger},
  {Hobbs}, {Klioner}, {Lammers}, {McMillan}, {Ramos-Lerate},
  {Steidelm{\"u}ller}, {Stephenson}, \& {van Leeuwen}}]{Lindegren20}
{Lindegren}, L., {Bastian}, U., {Biermann}, M., {et~al.} 2020, arXiv e-prints,
  arXiv:2012.01742.
\newblock \doarXiv{2012.01742}

\bibitem[{{Lynch} {et~al.}(2018){Lynch}, {Swiggum}, {Kondratiev}, {Kaplan},
  {Stovall}, {Fonseca}, {Roberts}, {Levin}, {DeCesar}, {Cui}, {Cenko},
  {Gatkine}, {Archibald}, {Banaszak}, {Biwer}, {Boyles}, {Chawla}, {Dartez},
  {Day}, {Ford}, {Flanigan}, {Hessels}, {Hinojosa}, {Jenet}, {Karako-Argaman},
  {Kaspi}, {Leake}, {Lunsford}, {Martinez}, {Mata}, {McLaughlin}, {Noori},
  {Ransom}, {Rohr}, {Siemens}, {Spiewak}, {Stairs}, {van Leeuwen}, {Walker}, \&
  {Wells}}]{Lynch18}
{Lynch}, R.~S., {Swiggum}, J.~K., {Kondratiev}, V.~I., {et~al.} 2018, \apj,
  859, 93, \dodoi{10.3847/1538-4357/aabf8a}

\bibitem[{{Miller} {et~al.}(2020){Miller}, {Swihart}, {Strader}, {Urquhart},
  {Aydi}, {Chomiuk}, {Dage}, {Kawash}, {Shishkovsky}, \&
  {Sokolovsky}}]{Miller20}
{Miller}, J.~M., {Swihart}, S.~J., {Strader}, J., {et~al.} 2020, \apj, 904, 49,
  \dodoi{10.3847/1538-4357/abbb2e}

\bibitem[{{Monet} {et~al.}(2003){Monet}, {Levine}, {Canzian}, {Ables}, {Bird},
  {Dahn}, {Guetter}, {Harris}, {Henden}, {Leggett}, {Levison}, {Luginbuhl},
  {Martini}, {Monet}, {Munn}, {Pier}, {Rhodes}, {Riepe}, {Sell}, {Stone},
  {Vrba}, {Walker}, {Westerhout}, {Brucato}, {Reid}, {Schoening}, {Hartley},
  {Read}, \& {Tritton}}]{Monet03}
{Monet}, D.~G., {Levine}, S.~E., {Canzian}, B., {et~al.} 2003, \aj, 125, 984,
  \dodoi{10.1086/345888}

\bibitem[{{Montes} {et~al.}(1995){Montes}, {Fernandez-Figueroa}, {de Castro},
  \& {Cornide}}]{Montes95}
{Montes}, D., {Fernandez-Figueroa}, M.~J., {de Castro}, E., \& {Cornide}, M.
  1995, \aap, 294, 165

\bibitem[{{Nolan} {et~al.}(2012){Nolan}, {Abdo}, {Ackermann}, {Ajello},
  {Allafort}, {Antolini}, {Atwood}, {Axelsson}, {Baldini}, {Ballet}, \&
  et~al.}]{2FGL}
{Nolan}, P.~L., {Abdo}, A.~A., {Ackermann}, M., {et~al.} 2012, \apjs, 199, 31,
  \dodoi{10.1088/0067-0049/199/2/31}

\bibitem[{{Paduano} {et~al.}(2021){Paduano}, {Bahramian}, {Miller-Jones},
  {Kawka}, {Strader}, {Chomiuk}, {Heinke}, {Maccarone}, {Britt}, {Plotkin},
  {Shaw}, {Shishkovsky}, {Tremou}, {Tudor}, \& {Sivakoff}}]{Paduano21}
{Paduano}, A., {Bahramian}, A., {Miller-Jones}, J.~C.~A., {et~al.} 2021

\bibitem[{{Pandey} \& {Singh}(2012)}]{Pandey12}
{Pandey}, J.~C., \& {Singh}, K.~P. 2012, \mnras, 419, 1219,
  \dodoi{10.1111/j.1365-2966.2011.19776.x}

\bibitem[{{Papitto} \& {de Martino}(2020)}]{Papitto20}
{Papitto}, A., \& {de Martino}, D. 2020, arXiv e-prints, arXiv:2010.09060.
\newblock \doarXiv{2010.09060}

\bibitem[{{Papitto} \& {Torres}(2015)}]{Papitto15}
{Papitto}, A., \& {Torres}, D.~F. 2015, \apj, 807, 33,
  \dodoi{10.1088/0004-637X/807/1/33}

\bibitem[{{Papitto} {et~al.}(2013){Papitto}, {Ferrigno}, {Bozzo}, {Rea},
  {Pavan}, {Burderi}, {Burgay}, {Campana}, {di Salvo}, {Falanga},
  {Filipovi{\'c}}, {Freire}, {Hessels}, {Possenti}, {Ransom}, {Riggio},
  {Romano}, {Sarkissian}, {Stairs}, {Stella}, {Torres}, {Wieringa}, \&
  {Wong}}]{Papitto13}
{Papitto}, A., {Ferrigno}, C., {Bozzo}, E., {et~al.} 2013, \nat, 501, 517,
  \dodoi{10.1038/nature12470}

\bibitem[{{Papitto} {et~al.}(2018){Papitto}, {Rea}, {Coti Zelati}, {de
  Martino}, {Scaringi}, {Campana}, {de O{\'n}a Wilhelmi}, {Knigge},
  {Serenelli}, {Stella}, {Torres}, {D'Avanzo}, \& {Israel}}]{Papitto18}
{Papitto}, A., {Rea}, N., {Coti Zelati}, F., {et~al.} 2018, \apjl, 858, L12,
  \dodoi{10.3847/2041-8213/aabee9}

\bibitem[{{R Core Team}(2017)}]{Rmanual}
{R Core Team}. 2017, R: A Language and Environment for Statistical Computing, R
  Foundation for Statistical Computing, Vienna, Austria

\bibitem[{{Ray} {et~al.}(2012){Ray}, {Abdo}, {Parent}, {Bhattacharya},
  {Bhattacharyya}, {Camilo}, {Cognard}, {Theureau}, {Ferrara}, {Harding},
  {Thompson}, {Freire}, {Guillemot}, {Gupta}, {Roy}, {Hessels}, {Johnston},
  {Keith}, {Shannon}, {Kerr}, {Michelson}, {Romani}, {Kramer}, {McLaughlin},
  {Ransom}, {Roberts}, {Saz Parkinson}, {Ziegler}, {Smith}, {Stappers},
  {Weltevrede}, \& {Wood}}]{Ray12}
{Ray}, P.~S., {Abdo}, A.~A., {Parent}, D., {et~al.} 2012, arXiv e-prints,
  arXiv:1205.3089.
\newblock \doarXiv{1205.3089}

\bibitem[{{Roberts}(2013)}]{Roberts13}
{Roberts}, M. S.~E. 2013, in IAU Symposium, Vol. 291, Neutron Stars and
  Pulsars: Challenges and Opportunities after 80 years, ed. J.~{van Leeuwen},
  127--132

\bibitem[{{Roberts} {et~al.}(2018){Roberts}, {Al Noori}, {Torres},
  {McLaughlin}, {Gentile}, {Hessels}, {Ransom}, {Ray}, {Kerr}, \&
  {Breton}}]{Roberts18}
{Roberts}, M. S.~E., {Al Noori}, H., {Torres}, R.~A., {et~al.} 2018, in Pulsar
  Astrophysics the Next Fifty Years, ed. P.~{Weltevrede}, B.~B.~P. {Perera},
  L.~L. {Preston}, \& S.~{Sanidas}, Vol. 337, 43--46

\bibitem[{{Rodono} {et~al.}(1995){Rodono}, {Lanza}, \& {Catalano}}]{Rodono95}
{Rodono}, M., {Lanza}, A.~F., \& {Catalano}, S. 1995, \aap, 301, 75

\bibitem[{{Roy} {et~al.}(2015){Roy}, {Ray}, {Bhattacharyya}, {Stappers},
  {Chengalur}, {Deneva}, {Camilo}, {Johnson}, {Wolff}, {Hessels}, {Bassa},
  {Keane}, {Ferrara}, {Harding}, \& {Wood}}]{Roy15}
{Roy}, J., {Ray}, P.~S., {Bhattacharyya}, B., {et~al.} 2015, \apjl, 800, L12,
  \dodoi{10.1088/2041-8205/800/1/L12}

\bibitem[{Ruf(1999)}]{Ruf99}
Ruf, T. 1999, Biological Rhythm Research, 30, 178

\bibitem[{{Schinzel} {et~al.}(2017){Schinzel}, {Petrov}, {Taylor}, \&
  {Edwards}}]{Schinzel17}
{Schinzel}, F.~K., {Petrov}, L., {Taylor}, G.~B., \& {Edwards}, P.~G. 2017,
  \apj, 838, 139, \dodoi{10.3847/1538-4357/aa6439}

\bibitem[{{Schlafly} \& {Finkbeiner}(2011)}]{Schlafly11}
{Schlafly}, E.~F., \& {Finkbeiner}, D.~P. 2011, \apj, 737, 103,
  \dodoi{10.1088/0004-637X/737/2/103}

\bibitem[{{Shappee} {et~al.}(2014){Shappee}, {Prieto}, {Grupe}, {Kochanek},
  {Stanek}, {De Rosa}, {Mathur}, {Zu}, {Peterson}, {Pogge}, {Komossa}, {Im},
  {Jencson}, {Holoien}, {Basu}, {Beacom}, {Szczygie{\l}}, {Brimacombe},
  {Adams}, {Campillay}, {Choi}, {Contreras}, {Dietrich}, {Dubberley},
  {Elphick}, {Foale}, {Giustini}, {Gonzalez}, {Hawkins}, {Howell}, {Hsiao},
  {Koss}, {Leighly}, {Morrell}, {Mudd}, {Mullins}, {Nugent}, {Parrent},
  {Phillips}, {Pojmanski}, {Rosing}, {Ross}, {Sand}, {Terndrup}, {Valenti},
  {Walker}, \& {Yoon}}]{Shappee14}
{Shappee}, B.~J., {Prieto}, J.~L., {Grupe}, D., {et~al.} 2014, \apj, 788, 48,
  \dodoi{10.1088/0004-637X/788/1/48}

\bibitem[{{Stacey} {et~al.}(2012){Stacey}, {Heinke}, {Cohn}, {Lugger}, \&
  {Bahramian}}]{Stacey12}
{Stacey}, W.~S., {Heinke}, C.~O., {Cohn}, H.~N., {Lugger}, P.~M., \&
  {Bahramian}, A. 2012, \apj, 751, 62, \dodoi{10.1088/0004-637X/751/1/62}

\bibitem[{{Stappers} {et~al.}(2014){Stappers}, {Archibald}, {Hessels}, {Bassa},
  {Bogdanov}, {Janssen}, {Kaspi}, {Lyne}, {Patruno}, {Tendulkar}, {Hill}, \&
  {Glanzman}}]{Stappers14}
{Stappers}, B.~W., {Archibald}, A.~M., {Hessels}, J.~W.~T., {et~al.} 2014,
  \apj, 790, 39, \dodoi{10.1088/0004-637X/790/1/39}

\bibitem[{{Strader} {et~al.}(2016){Strader}, {Li}, {Chomiuk}, {Heinke},
  {Udalski}, {Peacock}, {Shishkovsky}, \& {Tremou}}]{Strader16}
{Strader}, J., {Li}, K.-L., {Chomiuk}, L., {et~al.} 2016, \apj, 831, 89,
  \dodoi{10.3847/0004-637X/831/1/89}

\bibitem[{{Strader} {et~al.}(2019){Strader}, {Swihart}, {Chomiuk}, {Bahramian},
  {Britt}, {Cheung}, {Dage}, {Halpern}, {Li}, {Mignani}, {Orosz}, {Peacock},
  {Salinas}, {Shishkovsky}, \& {Tremou}}]{Strader19}
{Strader}, J., {Swihart}, S., {Chomiuk}, L., {et~al.} 2019, \apj, 872, 42,
  \dodoi{10.3847/1538-4357/aafbaa}

\bibitem[{{Stroh} \& {Falcone}(2013)}]{Stroh13}
{Stroh}, M.~C., \& {Falcone}, A.~D. 2013, \apjs, 207, 28,
  \dodoi{10.1088/0067-0049/207/2/28}

\bibitem[{{Swihart} {et~al.}(2018){Swihart}, {Strader}, {Shishkovsky},
  {Chomiuk}, {Bahramian}, {Heinke}, {Miller-Jones}, {Edwards}, \&
  {Cheung}}]{Swihart18}
{Swihart}, S.~J., {Strader}, J., {Shishkovsky}, L., {et~al.} 2018, \apj, 866,
  83, \dodoi{10.3847/1538-4357/aadcab}

\bibitem[{{Tauris} \& {Savonije}(1999)}]{Tauris99}
{Tauris}, T.~M., \& {Savonije}, G.~J. 1999, \aap, 350, 928.
\newblock \doarXiv{astro-ph/9909147}

\bibitem[{{Tendulkar} {et~al.}(2014){Tendulkar}, {Yang}, {An}, {Kaspi},
  {Archibald}, {Bassa}, {Bellm}, {Bogdanov}, {Harrison}, {Hessels}, {Janssen},
  {Lyne}, {Patruno}, {Stappers}, {Stern}, {Tomsick}, {Boggs}, {Chakrabarty},
  {Christensen}, {Craig}, {Hailey}, \& {Zhang}}]{Tendulkar14}
{Tendulkar}, S.~P., {Yang}, C., {An}, H., {et~al.} 2014, \apj, 791, 77,
  \dodoi{10.1088/0004-637X/791/2/77}

\bibitem[{{Tody}(1986)}]{Tody86}
{Tody}, D. 1986, in Society of Photo-Optical Instrumentation Engineers (SPIE)
  Conference Series, Vol. 627, Instrumentation in astronomy VI, ed. D.~L.
  {Crawford}, 733

\bibitem[{{Udalski} {et~al.}(2015){Udalski}, {Szyma{\'n}ski}, \&
  {Szyma{\'n}ski}}]{Udalski15}
{Udalski}, A., {Szyma{\'n}ski}, M.~K., \& {Szyma{\'n}ski}, G. 2015, \actaa, 65,
  1.
\newblock \doarXiv{1504.05966}

\bibitem[{{van Paradijs}(1998)}]{Paradijs98}
{van Paradijs}, J. 1998, in NATO Advanced Science Institutes (ASI) Series C,
  Vol. 515, NATO Advanced Science Institutes (ASI) Series C, ed. R.~{Buccheri},
  J.~{van Paradijs}, \& A.~{Alpar}, 279

\bibitem[{{Veledina} {et~al.}(2019){Veledina}, {N{\"a}ttil{\"a}}, \&
  {Beloborodov}}]{Veledina19}
{Veledina}, A., {N{\"a}ttil{\"a}}, J., \& {Beloborodov}, A.~M. 2019, \apj, 884,
  144, \dodoi{10.3847/1538-4357/ab44c6}

\bibitem[{{Wilms} {et~al.}(2000){Wilms}, {Allen}, \& {McCray}}]{Wilms00}
{Wilms}, J., {Allen}, A., \& {McCray}, R. 2000, \apj, 542, 914,
  \dodoi{10.1086/317016}

\end{thebibliography}

\begin{deluxetable*}{lrr}
\tablecaption{SOAR/Goodman Photometry of J0540B}

\tablehead{
\colhead{MBJD} &
\colhead{$G$} & 
\colhead{unc.} \\
\colhead{(d)} &
\colhead{(mag)} &
\colhead{(mag)} 
}
\startdata
59227.0364879 & 20.437 & 0.034 \\
59227.0389600 & 20.261 & 0.023 \\
59227.0393720 & 20.408 & 0.027 \\
59227.0397848 & 20.378 & 0.026 \\
59227.0401971 & 20.295 & 0.023 \\
59227.0406092 & 20.289 & 0.023 \\
59227.0410215 & 20.259 & 0.021 \\
59227.0414339 & 20.151 & 0.019 \\
59227.0418460 & 20.175 & 0.019 \\
59227.0422582 & 20.196 & 0.020 \\
59227.0426706 & 20.144 & 0.019 \\
59227.0430826 & 20.159 & 0.020 \\
\nodata & \nodata & \nodata 
\enddata
\tablecomments{Table 1 is published in its entirety in the machine-readable format. A portion is shown here for guidance regarding its form and content. All dates are Modified Barycentric Julian Dates (BJD--2400000.5) on the TDB system \citep{Eastman10}.}
\label{tab:soar_tab}
\end{deluxetable*}
 
\end{document}